%% file: samplesize.tex
\documentclass[twocolumn, times, 5p, number, sort&compress]{elsarticle}\usepackage{graphicx, xcolor}
\makeatletter
\def\maxwidth{ %
  \ifdim\Gin@nat@width>\linewidth
    \linewidth
  \else
    \Gin@nat@width
  \fi
}
\makeatother

\IfFileExists{upquote.sty}{\usepackage{upquote}}{}
\definecolor{fgcolor}{rgb}{0.2, 0.2, 0.2}
\newcommand{\hlfunctioncall}[1]{\textcolor[rgb]{0.501960784313725,0,0.329411764705882}{\textbf{#1}}}%
\newcommand{\hlstring}[1]{\textcolor[rgb]{0.6,0.6,1}{#1}}%

\usepackage{framed}
\makeatletter
\newenvironment{kframe}{%
 \def\at@end@of@kframe{}%
 \ifinner\ifhmode%
  \def\at@end@of@kframe{\end{minipage}}%
  \begin{minipage}{\columnwidth}%
 \fi\fi%
 \def\FrameCommand##1{\hskip\@totalleftmargin \hskip-\fboxsep
 \colorbox{shadecolor}{##1}\hskip-\fboxsep
     \hskip-\linewidth \hskip-\@totalleftmargin \hskip\columnwidth}%
 \MakeFramed {\advance\hsize-\width
   \@totalleftmargin\z@ \linewidth\hsize
   \@setminipage}}%
 {\par\unskip\endMakeFramed%
 \at@end@of@kframe}
\makeatother

\definecolor{shadecolor}{rgb}{.97, .97, .97}
\definecolor{messagecolor}{rgb}{0, 0, 0}
\definecolor{warningcolor}{rgb}{1, 0, 1}
\definecolor{errorcolor}{rgb}{1, 0, 0}
\newenvironment{knitrout}{}{} % an empty environment to be redefined in TeX

\usepackage{alltt}
\usepackage[utf8]{inputenx}
\usepackage[T1]{fontenc}
\usepackage{graphicx}
\usepackage[pdftex,
             unicode = true,
             bookmarks = true,
             bookmarksnumbered = false,
             bookmarksopen = false,
             breaklinks = false,
             backref = false,
             colorlinks = true,
             citecolor=black,        % color of links to bibli
             filecolor=black,      % color of file links
             urlcolor=black,           % color of external links
            linkcolor=black          % color of internal links
           ]{hyperref}
\usepackage{hypernat}
\usepackage{xspace}
\usepackage{mdwlist}
\usepackage{amsfonts}
\usepackage{amssymb}
\usepackage{marvosym}
\usepackage{textcomp}
\usepackage{pifont}
\usepackage{booktabs}
\usepackage{subfig}
\usepackage{array}
\usepackage[version=3]{mhchem}
\usepackage{xspace}
\usepackage{url}
\usepackage{fixltx2e}
\makeatletter
\def\url@cbstyle{%
  \@ifundefined{selectfont}{\def\UrlFont{}}{\def\UrlFont{\small}}}
\makeatother
\usepackage{tikz}
\usepackage{siunitx}
\colorlet{iphtblue}{black}
\colorlet{darkblue}{blue!50!black}
\usetikzlibrary{decorations.pathreplacing}%\usetikzlibrary{snakes}
\usetikzlibrary{matrix} 
\usetikzlibrary{patterns}

\newcommand{\ie}{\textit{i.\,e.}\xspace}

\newcommand{\eg}{\textit{e.\,g.}\xspace}
\newcommand{\Eg}{\textit{E.\,g.}\xspace}

\newcommand{\rcm}[1][\,]{#1cm\textsuperscript{-1}\xspace}
\newcommand{\tes}[1][th]{\textsuperscript{#1}\xspace}

\newcommand{\pow}[1]{\,$\cdot$\,10\textsuperscript{#1}\xspace}

\newcommand{\ntest}{\ensuremath{n_{test}}\xspace}
\newcommand{\ntrain}{\ensuremath{n_{train}}\xspace}

\usepackage{ccicons}

\DeclareMathOperator{\Var}{Var}
\DeclareMathOperator{\Sens}{Sens}
\DeclareMathOperator{\Spez}{Spec}
\DeclareMathOperator{\NPV}{NPV}
\DeclareMathOperator{\PPV}{PPV}

\newlength{\hlw}
\newcommand{\fancybox}[3][red]{
\begin{tikzpicture}
\node [draw=#1, very thick, rectangle, rounded corners, inner sep=10pt] (box){%
    \begin{minipage}{\linewidth-20pt}
      \vspace{0.5\baselineskip}
      #3
    \end{minipage}
};
\node[fill = #1, text = white, right=5mm, rounded corners] at (box.north west)
  {\sffamily\bfseries\large #2};
\end{tikzpicture}%
} 

\begin{document}
\begin{frontmatter}
  \title{Sample Size Planning for Classification Models} \author[ipht]{Claudia~Beleites\corref{cor}}
  \ead{Claudia.Beleites@ipht-jena.de} \cortext[cor]{Corresponding author}
  \author[ipht,cscc]{Ute~Neugebauer} \author[ipc]{Thomas~Bocklitz} \author[ipht]{Christoph~Krafft}
  \author[ipht,cscc,ipc]{Jürgen~Popp} \address[ipht]{Department of Spectroscopy and Imaging,
    Institute of Photonic Technology, Albert-Einstein-Str. 9, 07745 Jena, Germany}
  \address[cscc]{Center for Sepsis Control and Care, Jena University Hospital, Erlanger Allee 101,
    07747 Jena, Germany} \address[ipc]{Institute of Physical Chemistry and Abbé Center of Photonics,
    Friedrich-Schiller-University Jena, Helmholtzweg 4, 07743 Jena, Germany}
\begin{abstract}
  In biospectroscopy, suitably annotated and statistically independent samples (\eg patients,
  batches, etc.) for classifier training and testing are scarce and costly.  Learning curves show the
  model performance as function of the training sample size and can help to determine the sample size
  needed to train good classifiers. However, building a good model is actually not enough: the
  performance must also be proven.  We discuss learning curves for typical small sample size
  situations with 5 -- 25 independent samples per class. Although the classification models achieve
  acceptable performance, the learning curve can be completely masked by the random testing
  uncertainty due to the equally limited test sample size.  In consequence, we determine test sample
  sizes necessary to achieve reasonable precision in the validation and find that 75 -- 100 samples
  will usually be needed to test a good but not perfect classifier.  Such a data set will then allow
  refined sample size planning on the basis of the achieved performance.  We also demonstrate how to
  calculate necessary sample sizes in order to show the superiority of one classifier over another:
  this often requires hundreds of statistically independent test samples or is even theoretically
  impossible.  We demonstrate our findings with a data set of ca. 2550 Raman spectra of single cells
  (five classes: erythrocytes, leukocytes and three tumour cell lines BT-20, MCF-7 and OCI-AML3) 
  as well as by an extensive simulation that allows precise determination of the actual performance 
  of the models in question.
\end{abstract}
\end{frontmatter}

\noindent\fancybox[blue!50!black]{Accepted Author Manuscript}{This paper has been published as\\
  \href{http://dx.doi.org/10.1016/j.aca.2012.11.007}{C.~Beleites, U.~Neugebauer, T.~Bocklitz, C.~Krafft and J.~Popp: \emph{Sample size planning for classification models}. Analytica Chimica Acta, 2013, 760 (Special Issue: Chemometrics in Analytical Chemistry 2012), 25--33, DOI: 10.1016/j.aca.2012.11.007}.\\
  The manuscript is also available at \href{http://arxiv.org/abs/1211.1323}{arXiv no. 1211.1323}, where the source files contain also the source code shown in supplementary file II.}

\input{text.tex}

\bibliographystyle{elsarticle-num}
\bibliography{Literatur,R}

\appendix
\onecolumn

\input{supplementary-confmat.tex}

\input{supplementary-code.tex}
\end{document}

%% file: text.tex
\section{Introduction}
\label{sec:introduction}

Sample size planning is an important aspect in the design of experiments. While this study explicitly
targets sample size planning in the context of biospectroscopic classification, the ideas and
conclusions apply to a much wider range of applications. Biospectroscopy suffers from extreme
scarcity of statistically independent samples, but small sample size problems are common also in many
other fields of application.

In the context of biospectroscopic studies, suitably annotated and statistically independent samples
for classifier training and validation frequently are rare and costly.  Moreover, the classification
problems are often rather \emph{ill-posed} (\eg diseased \emph{vs.} non-diseased). In these
situations, particular classes are extremely rare, and/or large sample sizes are necessary to cover
classes that are rather ill-defined like ``not this disease'' or ``out of specification''. In
addition, ethical considerations often restrict the studied number of patients or animals.

Even though the data sets often consist of thousands of spectra, the statistically relevant number of
\emph{independent} cases is often extremely small due to ``hierarchical'' structure of the
biospectroscopic data sets: many spectra are taken of the same specimen, and possibly multiple
specimen of the same patient are available. Or, many spectra are taken of each cell, and a number of
cells is measured for each cultivation batch, etc. In these situations, the number of statistically
independent cases is given by the sample size on the highest level of the data hierarchy, \ie
patients or cell culture batches. All these reasons together lead to sample sizes that are typically
in the order of magnitude between 5 and 25 statistically independent cases per class.

Learning curves describe the development of the performance of chemometric models as function of the
training sample size. The true performance depends on the difficulty of the task at hand and must
therefore be measured by preliminary experiments. Estimation of necessary sample sizes for medical
classification has been done based on learning curves \cite{Mukherjee2003, Figueroa2012} as well as
on model based considerations \cite{Dobbin2007, Dobbin2008}. In pattern recognition, necessary
training sample sizes have been discussed for a long time (\eg \cite{Jain1982, Raudys1991,
  Kalayeh1983}).

However, building a good model is not enough: the quality of the model needs to be demonstrated.

One may think of training a classifier as the process of \emph{measuring} the model parameters
(coefficients etc.). Likewise, testing a classifier can be described as a \emph{measurement} of the
model performance. Like other measured values, both the parameters of the model and the observed
performance are subject to systematic (bias) and random (variance) uncertainty.

\newcommand{\zufrac}[5]{
  \small
  \matrix at (#1) (#2-zaehler) [matrix of nodes, cells = {draw}, 
  nodes = {draw, text width = 1ex, text height = 0.75ex, text depth = 0.25ex},
  execute at empty cell = {\node{};},
  row sep = -0.5pt,
  column sep = -0.5pt,
  ampersand replacement=\&,
  #3
]{#4};
\draw[very thick] ([below = 0.5ex] #2-zaehler.south west) ++(-.5ex, 0pt) node[coordinate] (#2-west) {}  -- 
                  ([below = 0.5ex] #2-zaehler.south east) -- ++(+.5ex, 0pt) node[coordinate] (#2-east) {};

 \matrix (#2-nenner) at (#2-zaehler.south)  [below = 1ex, matrix of nodes, cells = {draw}, 
  nodes = {draw, text width = 1ex, text height = 0.75ex, text depth = 0.25ex},
  execute at empty cell = {\node{};},
  row sep = -0.5pt,
  column sep = -0.5pt,
  ampersand replacement=\&
]{#5};
}
\newcommand{\tdiag}[1][0.75em]{\tikz[x=#1, y=#1, tight background]{
    \draw [ultra thin] (0, 0) rectangle (1, 1);  
    \draw (0, 0) -- (1, 1);
  }%
}
\newcommand{\tmdiag}[1][0.75em]{\tikz[x=#1, y=#1, tight background]{
    \draw [ultra thin] (0, 0) rectangle (1, 1);  
    \draw (0, 1) -- (1, 0);
  }%
}

\newcommand{\thv}[1][0.75em]{\tikz[x=#1, y=#1, tight background]{
    \draw [ultra thin] (0, 0) rectangle (1, 1);  
    \draw (0, .5) --++ (1, 0);
    \draw (.5, 0) --++ (0, 1);
  }%
}
\newcommand{\tp}[1][0.75em]{\tikz[x=#1, y=#1, tight background]{
    \draw [ultra thin] (0, 0) rectangle (1, 1);  
    \fill (0.5, .5) circle (#1/25);
  }%
}

\usetikzlibrary{backgrounds} 

\tikzstyle{thin}+=[line width=0.5pt]
\tikzstyle{shade}=[fill = gray]
\tikzstyle{eqmat}=[matrix of math nodes, below, 
  ampersand replacement=\&, column sep = 0pt,
  nodes = {anchor = base west, fill =yellow}]
 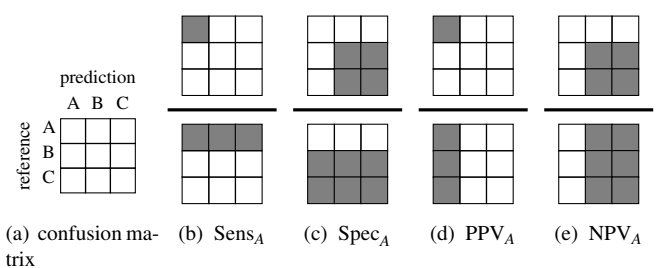
\begin{figure}
% \centering
\subfloat[\label{fig:Z} confusion matrix]{%
\begin{scriptsize}
\begin{tikzpicture}[tight background]
  \matrix (cm) [matrix of nodes, %cells = {draw}, 
  nodes = {text width = 1.5ex, text height = 1ex, text depth = 0.5ex, anchor = base east},
  execute at empty cell = {\node[draw]{};},
  row sep = -0.5pt,
  column sep = -0.5pt,
  ampersand replacement=\&,
  ]{
    ~ \& A \& B \& C \\
    A \&   \&   \&   \\
    B \&   \&   \&   \\
    C \&   \&   \&   \\
  };
  \node at (cm-1-3.north) [above, text width = 12ex, text badly centered] {prediction};
  \node at (cm-3-1.west) [xshift = -1.5ex, text width = 12ex, text badly centered, rotate = 90] 
  {reference} ;
\end{tikzpicture}
\end{scriptsize}}
\subfloat[\label{fig:sens}  $\Sens_A$]{
  \begin{tikzpicture}
    \zufrac{0,0}{senscl}{}
    {
      |[shade]| \&           \&            \\
                \&           \&            \\
                \&           \&            \\
    }{
      |[shade]| \& |[shade]| \& |[shade]|  \\
                \&           \&            \\
                \&           \&            \\
    }
 \end{tikzpicture}
}
\subfloat[\label{fig:spez} $\Spez_A$]{
  \begin{tikzpicture}
    \zufrac{0,0}{senscl}{}
    {
                \&           \&            \\
                \& |[shade]| \& |[shade]|  \\
                \& |[shade]| \& |[shade]|  \\
    }{
                \&           \&            \\
      |[shade]| \& |[shade]| \& |[shade]|  \\
      |[shade]| \& |[shade]| \& |[shade]|  \\
    }
   \end{tikzpicture}
}
\subfloat[\label{fig:ppv} $\PPV_A$]{
  \begin{tikzpicture}
    \zufrac{0,0}{senscl}{}
    {
      |[shade]| \&           \&            \\
                \&           \&            \\
                \&           \&            \\
    }{
      |[shade]| \&           \&            \\
      |[shade]| \&           \&            \\
      |[shade]| \&           \&            \\
    }
 \end{tikzpicture}
}
\subfloat[\label{fig:npv} $\NPV_A$]{
  \begin{tikzpicture}
    \zufrac{0,0}{senscl}{}
    {
                \&           \&            \\
                \& |[shade]| \& |[shade]|  \\
                \& |[shade]| \& |[shade]|  \\
    }{
                \& |[shade]| \& |[shade]|  \\
                \& |[shade]| \& |[shade]|  \\
                \& |[shade]| \& |[shade]|  \\
    }
   \end{tikzpicture}
}
\caption{ Confusion matrix \protect\subref{fig:Z} and characteristic fractions
  \protect\subref{fig:sens} -- \protect\subref{fig:npv}. The parts of the confusion matrix summed as
  enumerator and denominator for the respective fraction with respect to class $A$ are
  shaded.}
\label{fig:kenngr} 
\end{figure}
Classifier performance is often expressed in fractions of test cases, counted from different parts of
the confusion matrix, see fig.~\ref{fig:kenngr}.  These ratios summarize characteristic aspects of
performance like sensitivity (Sens$_A$: ``How well does the model recognize truly diseased
samples?'', fig.~\ref{fig:sens}), specificity (Spec$_A$: ``How well does the classifier recognize the
absence of the disease?'', fig.~\ref{fig:spez}), positive and negative predictive values
(PPV$_A$/NPV$_A$: ``Given the classifier diagnoses disease/non-disease, what is the probability that
this is true?'', fig.~\ref{fig:ppv} and \ref{fig:npv}). Sometimes further ratios, \eg the overall
fraction of correct predictions or misclassifications, are used.

The predictive values, while obviously of more interest to the user of a classifier than sensitivity
and specificity, cannot be calculated without knowing the relative frequencies (prior probabilities)
of the classes. 

From the sample size point of view, one important difference between these different ratios is the
number of test cases \ntest that appears in the denominator. This test sample size plays a crucial
role in determining the random uncertainty of the observed performance $\hat p$, (see
below). Particularly in multi-class problems, this test sample size varies widely: the number of test
cases truly belonging to the different classes may differ, leading to different and rather small test
sample sizes for determining the sensitivity $p$ of the different classes. On contrast, the overall
fraction of correct or misclassified samples use all tested samples in the denominator.

The specificity is calculated from all samples that truly do \emph{not} belong to the particular
class (fig.~\ref{fig:spez}). Compared to the sensitivities, the test sample size in the denominator
of the specificities is therefore usually larger and the performance estimate more precise (with the
exception of binary classification, where the specificity of one class is the sensitivity of the
other). Thus small sample size problems in the context of measuring classifier performance are better
illustrated with sensitivities. It should also be kept in mind that the specificity often corresponds
to an ill-posed question: ``\emph{Not class A}'' may be anything. Yet not all possibilities of a
sample truly not belonging to class A are of the same interest. In multi-class set-ups, the
specificity will often pool easy distinctions with more difficult differential diagnoses. In our
application \cite{Neugebauer2010, Neugebauer2010a}, the specificity for recognizing a cell does not
come from the BT-20 cell line pools \eg the fact that it is not an erythrocyte (which can easily be
determined by eye without any need for chemometric analysis) with the fact that it does not come from
the MCF-7 cell line, which is far more similar (yet from a clinical point of view possibly of low
interest as both are breast cancer cell lines) and the clinically important fact that it does not
belong to the OCI-AML3 leukemia. This pooling of all other classes has important
consequences. Increasing numbers of test cases in easily distinguished classes (erythrocytes) will
lead to improved specificities without any improvement for the clinically relevant differential
diagnoses. Also, it must be kept in mind that random predictions (guessing) already lead to
specificities that seem to be very good. For our real data set with five different classes, guessing
yields specificities between 0.77 and 0.85. Reported sensitivities should also be read in relation to
guessing performance, but neglecting to do so will not cause an intuitive overestimation of the
prediction quality: guessing sensitivities are around 0.20 in our five-class problem.

Examining the non-diagonal parts of the confusion table instead of specificities avoids these
problems. If reported as fractions of test cases truly belonging to that class, then all elements of
the confusion table behave like the sensitivities on the diagonal, if reported as fractions of cases
predicted to belong to that class, the entries behave like the positive predictive values (again on
the diagonal).

Literature guidance on how to obtain low total uncertainty and how to validate different aspects of
model performance is available \cite{Hastie2009, Dougherty2010, Kohavi1995, Beleites2005,
  Esbensen2010}.  In classifier testing, usually several assumptions are implicitly made which are
closely related to the behaviour of the performance measurements in terms of systematic and
random uncertainty.

Classification tests are usually described as Bernoulli-process (repeated coin throwing, following
a binomial distribution): \ntest samples are tested, and thereof $k$ successes (or errors) are
observed.  The true performance of the model is $p$, and its point estimate is
\begin{equation}
  \label{eq:hatp}
 \hat p = \frac{k}{\ntest} 
\end{equation}
with variance
\begin{equation}
  \label{eq:1}
 \Var \left(\frac{k}{\ntest}\right) = \frac{p (1 - p)}{\ntest} 
\end{equation}

In small sample size situations, resampling strategies like the bootstrap or repeated/iterated
$k$-fold cross validation are most appropriate. These strategies estimate the performance by setting
aside a (small) part of the samples for independent testing and building a model without these
samples, the \emph{surrogate model}. The surrogate model is then tested with the remaining
samples. The test results are refined by repeating/iterating this procedure a number of
times. Usually, the average performance over all surrogate models is reported. This is an
\emph{unbiased} estimate of the performance of models with the same training sample size as the
surrogate models \cite{Dougherty2010, Mukherjee2003}.  Note that the observed variance over the
surrogate models possibly underestimates the true variance of the performance of models trained with
\ntrain training cases \cite{Mukherjee2003}. This is intuitively clear if one thinks of a situation
where the surrogate models are perfectly stable, \ie different surrogate models yield the same
prediction for any given case. No variance is observed between different iterations of a $k$-fold
cross validation. Yet, the observed performance is still subject to the random uncertainty due to the
finite test sample size of the underlying Bernoulli process.

Usually, the performance measured with the surrogate models is used as approximation of the
performance of a model trained with all samples, the \emph{final} model. The underlying assumption is
that setting aside of the surrogate test data does not affect the model performance. In other
words, the learning curve is assumed to be flat between the training sample size of the surrogate
model and training sample size of the final model. The violation of this assumption causes the
well-known pessimistic bias of resampling based validation schemes. 

The results of testing many surrogate models are usually pooled.  Strictly speaking, pooling is
allowed only if the distributions of the pooled variables are equal. The description of the testing
procedure as Bernoulli process allows pooling if the surrogate models have equal true performance
$p$. In other words, if the predictions of the models are stable with respect to perturbed training
sets, \ie if exchanging of a few samples does not lead to changes in the prediction. Consequently,
model instability causes additional variance in the measured performance.

Here, we discuss the implications of these two aspects of sample size planning with a
Raman-spectroscopic five-class classification problem: the recognition of five different cell types
that can be present in blood. In addition to the measured data set, the results are complemented by a
simulation which allows arbitrary test precision.

\section{Materials and Methods}
\label{sec:materials-methods}
\begin{table*}[t]
    \begin{tabular}{llrrrrr}
    \toprule
    class & cell type          & n\textsubscript{spectra} & \multicolumn{4}{c}{sensitivity}                                              \\
          &                    &                          & sim. LDA & sim. PLS-LDA & real PLS-LDA        & real PLS-LDA batch-wise      \\
    \cmidrule(r){1-1} \cmidrule(lr){2-2}  \cmidrule(lr){3-3} \cmidrule(lr){4-4} \cmidrule(lr){5-5} \cmidrule(lr){6-6}  \cmidrule(l){7-7} \\ 
    rbc   & erythrocytes       & 372                      & 1.00     & 1.00         & 0.99 (0.96 -- 0.99) & 0.97 (0.96 -- 0.98)          \\
    leu   & leukocytes         & 569                      & 1.00     & 0.99         & 0.97 (0.96 -- 0.97) & 0.87 (0.84 -- 0.90)          \\
    mcf   & MCF-7 breast carc. & 558                      & 0.95     & 0.87         & 0.91 (0.90 -- 0.92) & 0.31 (0.24 -- 0.42)          \\
    bt    & BT 20 breast carc. & 532                      & 0.91     & 0.72         & 0.75 (0.74 -- 0.76) & 0.38 (0.32 -- 0.45)          \\
    oci   & OCI-AML3 leukemia  & 518                      & 0.94     & 0.86         & 0.89 (0.88 -- 0.90) & 0.30 (0.23 -- 0.17)          \\
    \bottomrule
  \end{tabular}
  \caption{Data set characteristics: classes, number of spectra per class and ``best possible'' sensitivities. For the simulated (sim.)  data 
    (column ``sim. LDA'' and ``sim. PLS-LDA''), \ntest = 2\pow4 spectra. Best possible performance of the real data was estimated using 100$\times$ 5-fold cross 
    validation, shown are average and 5\textsuperscript{th} to 95\textsuperscript{th} percentile of observed sensitivities over the 
    iterations. Column ``real PLS-LDA'' corresponds to the setup for this study, treating each spectrum as independent of the other spectra, for 
    column 7 (``real PLS-LDA batch-wise'')  the validation splits patients and batches rather than spectra.}
  \label{tab:characteristics}
\end{table*}

\subsection{Raman Spectra of Single Cells}
\label{sec:raman-spectra-single}

Raman spectra of five different types of cells that could be present in blood are used in this
study. Details of the preparation, measurements and the application have been published previously
\cite{Neugebauer2010, Neugebauer2010a}. The data were measured in a stratified manner, specifying
roughly equal numbers of cells per class beforehand, and do not reflect relative frequencies of the
different cells in a target patient population. Thus, we cannot calculate predictive values for our
classifiers.

For this study, the spectra were imported into R \cite{R} using package hyperSpec
\cite{hyperSpec}. In order to correct for deviations of the wavenumber calibration the maximum of the
\ce{CaF2} band was aligned to 322\rcm.  The spectra then underwent a smoothing interpolation
(\texttt{spc.loess}) onto a common wavenumber axis ranging from 500 to 1800 and 2600 to 3200\rcm with
data point spacing of 4\rcm.  Baseline correction was performed in the high wavenumber region by a
third order polynomial fit to spectral regions where no CH stretching signals occur (2700 -- 2825,
3020 -- 3040 and 3085 -- 3200 \rcm) which was then used as baseline for the CH stretching bands from
2810 to 3085 \rcm. A third order polynomial automatically selecting support points between 500 --
1200 \rcm was blended smoothly with a quadratic polynomial in the spectral range automatically
selecting support points between 800 -- 1200 and 1700 -- 1800 \rcm.  After baseline correction, the
spectral ranges 600 -- 1800 and 2810 -- 3085 \rcm were retained. Finally, the spectra were area
normalized.

\begin{figure}[t]
  \includegraphics[width=\hlw]{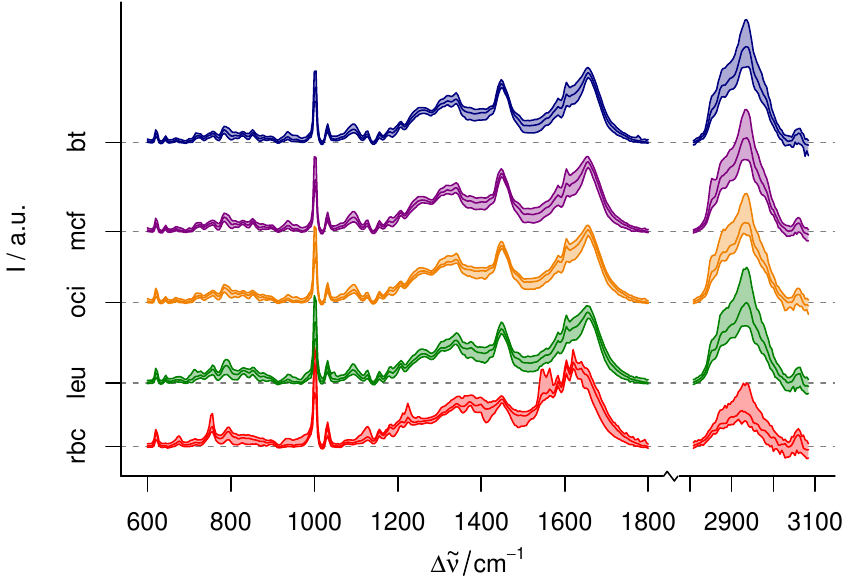}
  \caption{Spectra of the 5 classes: BT-20 breast carcinoma cells, MCF-7 breast carcinoma cells, OCI-AML3
    leukemia cells, normal leukocytes and normal erythrocytes (from top to bottom). Shown are the median
    and the 5\textsuperscript{th} to 95\textsuperscript{th} percentile spectra. The confusion tables
    are available as supplementary material.}
  \label{fig:spc}
\end{figure}
Figure~\ref{fig:spc} shows the preprocessed spectra. Erythrocyte (red blood cells, rbc) spectra can
easily be recognized by the resonance enhanced characteristic signature of hemoglobin around 1600
\rcm. Leukocyte (leu) spectra are rather similar to the tumour cell spectra, yet there are subtle
differences in the shape of the \ce{CH2}-deformation vibrations around 1440 \rcm, the intensity of
the $\nu_{CH}$ stretching vibrations (2810 -- 3085 \rcm) which are more intense in the tumour cells,
and the intensity of the phenylalanine band at 1002 \rcm (less intense in the tumour cells). Between
the different tumour cell lines (bt, mcf, and oci) no distinct marker bands are visible by eye.
 
Variation in the data set is introduced by using cells from 5 different donors (leukocytes and
erythrocytes) and 5 different cultivation batches, respectively; measuring the cells on the first day
of preparation and one day after (yielding 9 measurement days) and using two different lasers of the
same model from the same manufacturer. For the present study, we \emph{pretend not to know of these
  influencing factors} and treat the spectra as independent. This allows us to pretend that we have a
sufficiently large data set to run reference calculations that can be used as ground truth.  The
consequence is that no performance for the recognition of the cell lines in general can be inferred
from this study: the results would be heavily overoptimistic (tab.~\ref{tab:characteristics}, see
also \cite{Esbensen2010} for a discussion of representative testing).

Hence, we have a data set of about 2500 spectra (tab.~\ref{tab:characteristics}) of five classes with
``unknown'' influencing factors.  The difficulty in recognising the five different classes varies
widely: while erythrocytes are extemely easy to recognize, we expect that perfect recognition of
leukocytes is possible as well though we expect that more training cases are needed to achieve
this. Differential diagnosis of the cancer cell lines is more difficult, and substantial overlap
between the two breast carcinoma cell lines BT-20 and MCF-7 has been observed in previous studies
\cite{Neugebauer2010, Neugebauer2010a}. Throughout this paper, we discuss the sensitivities for
erythrocytes (rbc), leukocytes (leu) and the tumour cell line BT-20 (bt). 

Of these 2500 spectra, we draw data sets of size 25 cases\,/\,class keeping the remaining spectra
as a large test set to get a more precise estimate of the performance of the respective models.

rbc is the smallest class, its sensitivity can be estimated with a precision better than $\pm$~%
0.052 %
(95\,\% confidence interval at sensitivity of 0.5).

\subsection{Simulated Spectra}
\label{sec:simulated-spectra}

In addition to the experimental data set, simulations were used. This allows to study an idealized
situation: arbitrarily large test sets allow to measure the true performance with negligible random
uncertainty due to the testing. Thus, the random uncertainty due to model instability can be measured
with the simulations while these two sources of random uncertainty cannot be separated for the real
data.

For each of the five classes in the experimental data set, average spectrum and covariance matrix
were calculated. Multivariate normally distributed simulated spectra were simulated using
\texttt{rmvnorm} \cite{mvtnorm.1, mvtnorm.2}. Briefly, the Mersenne-Twister algorithm generates
uniformly distributed pseudo-random numbers which are then converted to normally distributed random
numbers via the inverse cumulative distribution function. The requested covariance structure is
obtained by multiplying with the matrix root of the covariance matrix (calculated via eigenvalue
decomposition) and the requested mean spectrum is added.

100 ``small'' data sets of 25 spectra\,/\,class (\ie 125 spectra of all classes together per small dataset) were generated. For
determining the real performance of the models, a large test set of 4\pow4 spectra\,/\,class was
generated. This means that the sensitivities can be measured with a precision of better than
0.5~$\pm$~%
0.005 %
(95\,\% c.i.), the standard deviation of observed performance is then
$\sigma (\hat p) = \sqrt{\frac{p (1 - p)}{n}} \leq \frac{0.5}{\sqrt{n}} = 0.0025$.

In addition, one large training set of 2\pow4 spectra\,/\,class was generated. This data set was
used to estimate the best possible performance that can be obtained with the chosen classifiers on
this idealized problem.

\subsection{Classification Models}
\label{sec:class-models}

As classifier we chose PLS-LDA as implemented in package \texttt{cbmodels}\cite{cbmodels} where the
partial least squares (PLS) and linear discriminant analysis (LDA) models from packages pls\cite{pls}
and MASS\cite{MASS} are combined into one model. The projection by the PLS is a suitable variable
reduction for LDA \cite{Barker2003}. LDA models trained on the PLS scores suffer much less from
instability than LDA models trained on data with large numbers of variates. The number of latent
variables was set to 10 for \ntrain $\geq$ 4 training spectra\,/\,class. For the extremely small
training sets, it was restricted to be at most half the total number of spectra in the training
set. All classification models were trained with all five classes.

In addition, we built two models using 2\pow4 simulated spectra\,/\,class and tested them with the large test set (4\pow4 spectra\,/\,class). These models are assumed
to achieve the best possible performance LDA can reach with and without PLS
dimensionality reduction for the given problem. The achieved sensitivities are 1.00 for rbc and
leu and 0.91 for bt (column ``sim. LDA'' in tab.~\ref{tab:characteristics}) without PLS. The 10
latent variable PLS-LDA model trained on the same data set had lower sensitivities of 1.00 for the
rbc, 0.99 for leu, and 0.72 for class bt (column ``sim PLS-LDA'').

For the real data, we report best possible performance for PLS-LDA models of the complete data set using 10 latent variables (measured by 100$\times$ iterated 5-fold cross validation, column ``real
PLS-LDA''). In addition, we checked the performance for 100$\times$ iterated 5-fold cross validation
when the validation splits are done by patient/batch (as the underlying structure of the measurement
would require; column ``real PLS-LDA batch-wise''). Here, 10 latent variable PLS-LDA can still
perfectly recognize erythrocytes, sensitivities for leukocytes are close to 0.90, but among the
tumour cell lines the model is basically guessing. 10 latent variable PLS-LDA is an extremely
restrictive model set-up which is appropriate for the small sample sizes studied in this paper but
recognition of circulating tumour cells requires more elaborate modelling \cite{Neugebauer2010,
  Neugebauer2010a}.

The interested reader will find the confusion tables, \ie sensitivities as well as the specificities
for the various types of misclassification,
in the supplementary material.

\subsection{Validation Set-Up}
\label{sec:validation-set-up}

Iterated $k$-fold cross validation was chosen as validation scheme. While out-of-bootstrap validation
is sometimes preferred for small sample sizes due to the lower variance, a previous study on
spectroscopic data sets found comparable overall uncertainty for these two validation schemes
\cite{Beleites2005}.  In contrast to $k$-fold cross validation, the effective training sample size is
not known in out-of-bootstrap validation. Out-of-bootstrap usually has the same nominal training
sample size as the whole data set. However, it is pessimistically biased with respect to the final
model. Such a pessimistic bias is usually observed if the training set is smaller than the whole data
set. This pessimistic bias is usually larger than that of 5- or 10-fold cross validation. This
suggests that the duplicate cases in the bootstrap training sets do not contribute as much
information for classifier training as the first instance of the given case does. Cross validation is
unbiased with respect to the number of cases actually used for training of the surrogate models
\cite{Dougherty2010} and is therefore more suitable for calculating learning curves.

We used $k = 5$-fold cross validation with 100 iterations.

\subsection{Growing Data Sets or Retrospective Learning Curves}
\label{sec:growing-data-sets}

Both real and simulated data sets were used for the learning curve estimation in a ``growing''
fashion. This simulates a scenario where at first very few cases are available, and new, better
models are built as further cases become available, following the practice of modeling and sample
collection we usually encounter. 

100 such growing data sets were analysed for both the real and the simulated data. This allows
calculation of the average performance that can be expected for our cell classifier with 10 latent
variable PLS-LDA models as well as the respective random uncertainty.

The alternative to the growing data set scenario, retrospective calculation of the learning curve,
would lead to an intermediate between the two different learning curves: as there are many
possibilities to draw few cases out of even a small data set, for the very small sample sizes
the resulting curve will be closer to the average performance of that training sample size. However, as
the drawn number of samples approaches the size of the small data set, the retrospective estimate of
the learning curve tends towards the estimate of the growing data set.
\section {Learning Curves}
\label{sec:learn-curv}
\begin{figure*}[ht!]
  \includegraphics[width=\linewidth]{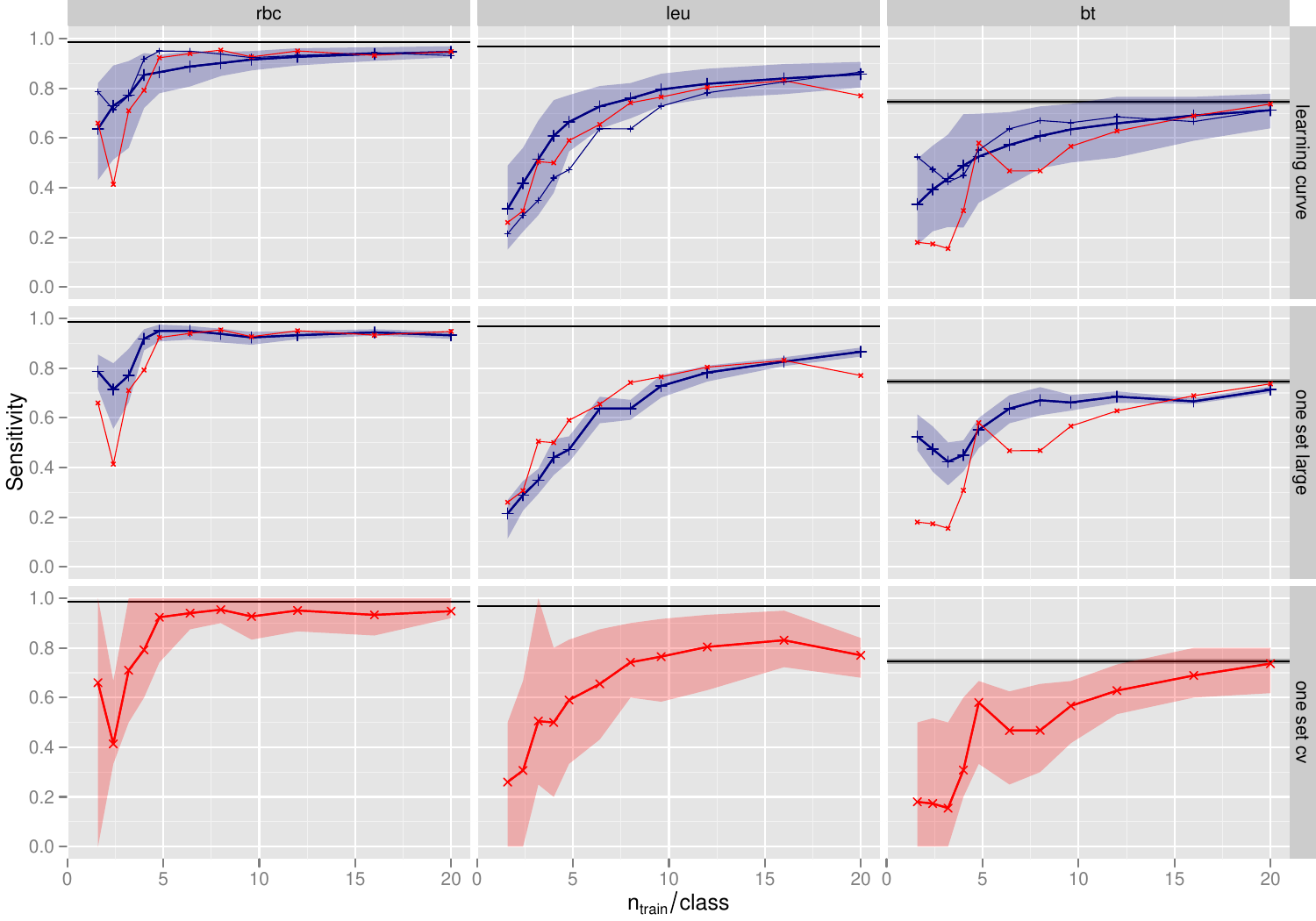}
  \caption{Learning curves of the real data set: sensitivities for recognition of red blood cells
    (rbc), leukocytes (leu), and BT-20 breast tumour cell line (bt).  
    Black: Sensitivity observed for 100 iterations of 5-fold cross validation on the complete data 
    set, approximating the best possible performance of a 10 latent variable PLS-LDA on this data 
    set. 
    Lines give the average, the shaded area covers the 5\tes to 95\tes percentile of iterations (bottom and middle row) and small data sets (top row). Thin lines: average ``one set large'' and ``one set cv'' (cross validation)  performance are repeated in the rows
    above for easier comparison. 
    Colours: {\color{darkblue}blue} performance measured with large test set, {\color{red}red}
    performance measured by iterated cross validation. 
    Bottom row: Learning curve of \emph{one} growing data set, measured with 100$\times$ iterated 
    5-fold cross validation.
    Middle row: The same models as in the bottom row, but performance measured with large test set. 
    The percentiles depict the instability of the surrogate models trained during iterated cross 
    validation, but are subject only to low uncertainty due to the finite \emph{test} sample size.
    Top row: sensitivity achieved for 100 different small data sets of size \ntrain, measured with the
    large test set.  
}
  \label{fig:real}
\end{figure*}
\begin{figure*}[tb]
  \includegraphics[width=\linewidth]{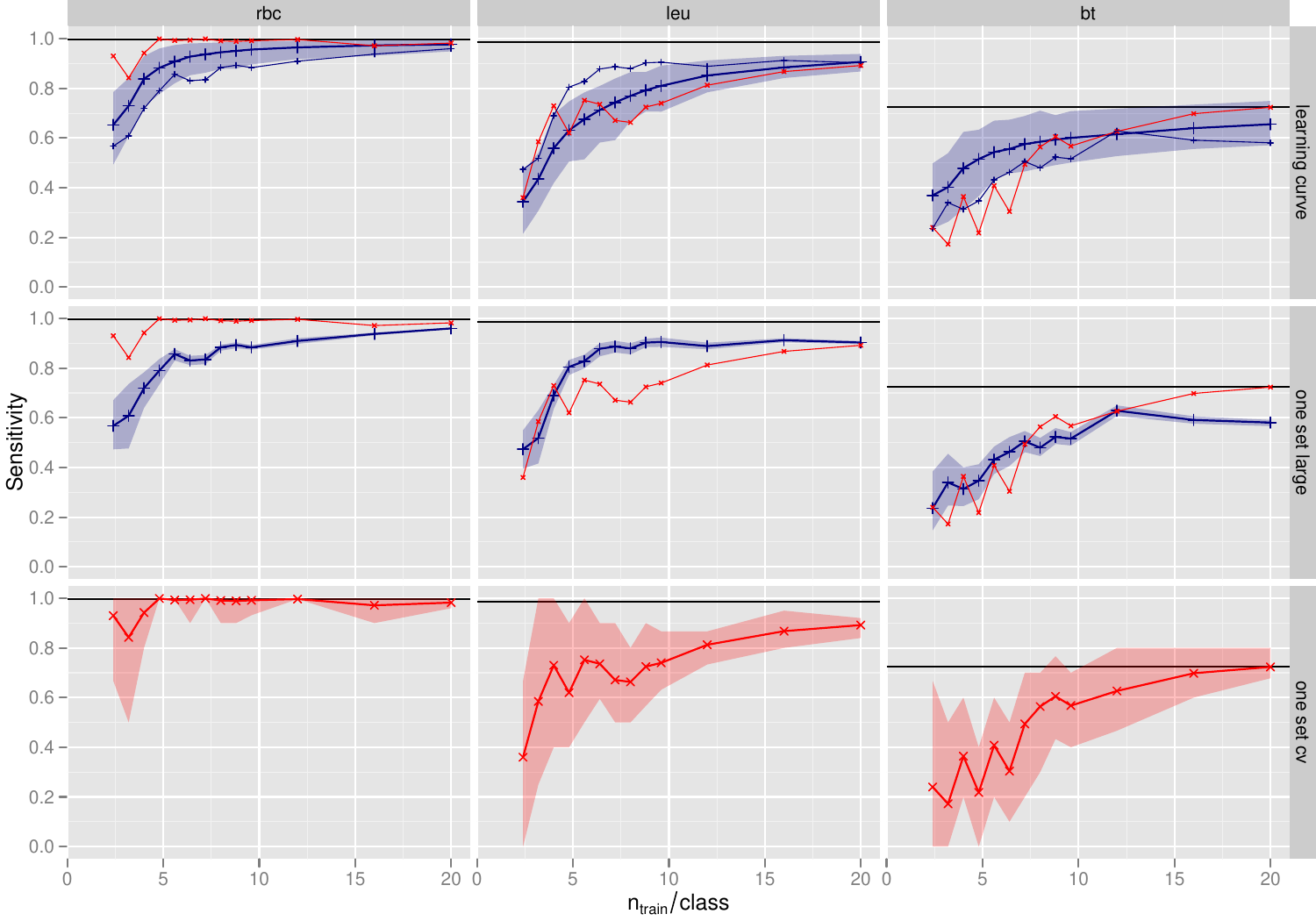}
  \caption{Learning curves of the simulated data set. This plot was generated analogously to
    fig~\ref{fig:real}. The only difference is that the best possible error (black lines) was measured with the
    large independent test set, see description of the data analysis.}
  \label{fig:sim}
\end{figure*}

The \emph{learning curve} describes the performance of a given classifier for a problem as function
of the training sample size \cite{Hastie2009}. The prediction errors a
classifier makes may be divided into four categories:
\begin{enumerate}
\item the irreducible or Bayes error
\item the bias due to the model setup,
\item additional systematic deviations (bias), and
\item random deviations (variance) 
\end{enumerate}
The best possible performance that can be achieved with a given model setup consists of the Bayes
error, \ie the best possible performance for the best possible model, and the bias for $\ntrain
\rightarrow \infty$.  The latter two components depend on the training sample size, and tend to zero
as more cases become available.

The general discussion of learning curves, \eg \cite[fig. 7.8]{Hastie2009},
usually considers the combination of the first three error types (as function of the training sample
size) which form the average (expected) performance of a given classification rule for a particular problem if
\ntrain training cases are available. The learning curve for a particular data set is known as
conditional learning curve \cite{Hastie2009}.

In the context of classification based on microarray data, both empirically fitted functions
\cite{Mukherjee2003} and parametric methods based on the difference in gene expression
\cite{Dobbin2007, Dobbin2008} have been used to estimate learning curves and necessary sample sizes
for the \emph{training} of well performing classifiers. An extension of Mukherjee \emph{et\,al.}
\cite{Mukherjee2003} has been applied to medical text classification \cite{Figueroa2012}.

Microarray (gene expression) data sets are similar to biospectroscopic data sets in their shape and
size: in both cases the raw data typically consists of thousands of measurement channels (variates:
genes, wavelengths) and typically hundreds to thousands of rows (expression profiles, spectra).
However, they differ from typical biospectroscopic data sets in two important aspects. Firstly,
biospectroscopic data sets often have rather large numbers of spectra of the same patient or batch
while multiple measurements of the same subject are far less common in microarray studies. The data
sets in Mukherjee \emph{et\,al.}  \cite{Mukherjee2003} have total patient numbers between 53 and 78
(plus one large set of 280 patients), these sample sizes unfortunately do not allow to check their
extrapolated predictions of the performance.
Secondly, the information with respect to the classification problem is usually spread out over wide
spectral ranges in biospectroscopic classification. In contrast, microarray classification typically
relies on rather few genes that carry information among a large number of noise-only variates
\cite{Dobbin2007, Dobbin2008}.

Figures~\ref{fig:real} and \ref{fig:sim} give the (unconditional) learning curves for the real and
simulated data in the top rows (lines). With smaller sample sizes, the random uncertainty grows, and
cannot be neglected: A \emph{particular} data set of size \ntrain may differ substantially from the
average data set of size \ntrain. For each training sample size, 90 of the 100 small data sets had
performance inside the shaded area.  

For the simulated data, one such growing data set is shown exemplarily in the middle (true
performance, \ie tested with the large test set) and bottom rows (cross validation estimate of
performance of the same model) of fig.~\ref{fig:sim}.  The example run performs exceptionally well
for the leukocytes but roughly at the 5\tes percentile with respect to all possible data sets of size
\ntrain of sensitivity for red blood cells and the BT-20 cell line. The example run of the real data
(fig.~\ref{fig:real}) in general follows more closely the average sensitivity of data sets of the
respective size.  Learning curves reported for real data sets usually give one point measurement for
each classifier set up and training sample size only and are usually calculated in the
``retrospective'' manner according to our definition above.

For the planning of necessary sample sizes needed to train good classifiers, both the expected
performance in the top row of figs.~\ref{fig:real} and \ref{fig:sim} and the performance for a given
growing data set as in the middle rows are of importance. The top rows answer the question how many
samples should be collected if no samples are yet available for a specific problem, while the middle
rows belong to the question how many \emph{more} samples in addition to the already available ones
should be collected. 

In practice, however, neither the top nor the middle row learning curves are available, only
(iterated) cross validation or out-of-bootstrap results are available from within a given data set.
The results of the cross-validation
in the bottom row are an unbiased estimate of the middle row (we use the actual training sample size
of the surrogate models, \ie $\frac{4}{5}$ of the sample size of the small data set). However, the
cross validation is subject to much higher random uncertainty, as the total number of test cases
is much lower than with the large test set used to calculate the middle rows. 

As explained before, the random uncertainty comes from two sources: firstly, model instability, \ie
differences between surrogate models built with different training sets of the same size, and
secondly testing uncertainty due to the finite number of spectra available for testing.  The first is
related to the number of \emph{training} samples while the second depends on the number of
\emph{test} samples. Testing with the large test set reduces the second source of uncertainty but
does not influence the variation due to model instability. The only difference between middle and
bottom rows in figs.~\ref{fig:real} and \ref{fig:sim} are the test sets: exactly the same models are
tested with the large test set (middle) and the spectra held out by the cross validation (bottom
row). In other words, the bottom row is a ``small \emph{test} sample size'' approximation to the
middle row. The simulations use \ntest = 2\pow4 for reference (top and middle row), meaning that the
variation depicted in middle row of fig.~\ref{fig:sim} is caused only by model instability. On
contrast, for the real data, only ca. 350 -- 540 reference test spectra are available and uncertainty
due to the finite test sample size can contribute substantially to the observed variation in the
middle row of fig.~\ref{fig:real}. However, the total random uncertainty on the iterated cross
validation is dominated by the huge random uncertainty due to testing only with the up to 25 samples
of the small data set.

This uncertainty is large enough to mask important features of the learning curve of the growing data
set: in our example run for the simulated data, the sensitivity for erythrocytes is largely
overestimated (other runs show equally large underestimation). The exceptionally good performance for
the leukocytes with 4 -- 10 training samples is not only not detected by the cross validation but in
fact two dips appear in the cross validation estimate of the example data set's learning curve. For
the BT-20 cell line, we observe an oscillating behaviour with the addition of single cases up to a
data set size of 9 samples (\ie on average 7.2 training samples). Of course, we observe also runs
that match the true (reference) learning curve of the particular data set more closely. But even then
the percentiles indicate that the results are not reliable estimates of the learning curve of that
data set.

The cross validation of the real data set underestimates the sensitivity for red blood cells for the
extremely small sample sizes, however the general development of sensitivity as function of the
training sample size of the example run is correctly reproduced. Also the learning curve for the
leukocytes is quite closely matched. For the BT-20 cells, however, the cross validation again does
not even resemble the shape of the example data set's learning curve.

In conclusion, the average performances observed during the iterated cross validation do not reliably
recover the correct shape of the learning curve of the particular data set for our small sample size
scenarios (middle rows), much less that of the performance of \emph{any} data set of the respective
training sample size (top rows). In contrast, the actual performance of the classifiers (top and
middle rows) is acceptable to very good considering the actual training sample sizes: with 20
training cases per class, red blood cells are almost perfectly recognized, sensitivities around
0.90 are achieved for leukocytes and even about 2 out of 3 of the very difficult BT-20 breast
cancer cells are recognized correctly.

\section{Sample Size Requirements for Classifier Testing}
Thus, the precise measurement of the classifier performance turns out to be more complicated in such
small sample size situations. Sample size planning for classification therefore needs to take into
account also the sample size requirements for the testing of the classifier. We will discuss here two
important scenarios that allow estimating required test sample sizes: firstly, specifying an
acceptable width for a confidence interval of the performance measure and secondly the number of test
cases needed for a comparison of classifiers.

\subsection{Specifying Acceptable Confidence Interval Widths}
\begin{knitrout}
\definecolor{shadecolor}{rgb}{0.969, 0.969, 0.969}\color{fgcolor}\begin{kframe}

{\ttfamily\noindent\itshape\textcolor{messagecolor}{\#\# Loading required package: ggplot2}}\end{kframe}
\end{knitrout}

For Bernoulli processes, several approaches exist to estimate confidence intervals for the true
probability $p$ given the observed probability $\hat p$ and the number of tests $n$, see
\cite{Brown2001, Pires2008} for recommendations particularly in small sample size situations.  For
the following discussion, we use the Bayes method with a uniform prior to obtain the minimal-length
or highest posterior density (HPD) interval \cite{binom, JaynesProbabilityTheory}. For details about
the statistical properties of this method, please refer to \cite{Pires2008}. Package
binom\cite{binom} offers a variety of other methods that can easily be used by the interested reader
instead. 

From a computational point of view, this method is convenient as the calculations can be formulated
using the Beta-distribution which allows to compute results not only for discrete numbers of events
$k$, but for real $k$. Thus, $\hat p$ obtained from testing many spectra can be used with a test
sample size \ntest equalling \eg the number of test \emph{patients} or \emph{batches}.

Confidence intervals for the true proportion are calculated as function of the number of test samples
(denominator of the proportion) and the observed proportion $\hat p$. The intervals are widest for
$\hat p = 0.5$ and narrowest for $\hat p = 0$ or $1$. Consequently, the necessary test sample size to
measure the performance with a pre-specified precision can be calculated, either in a conservative
(worst-case) fashion for $\hat p = 0.5$ or using existing knowledge/expectations about the achievable
performance.

\begin{figure}[tb]
\includegraphics[width=\hlw]{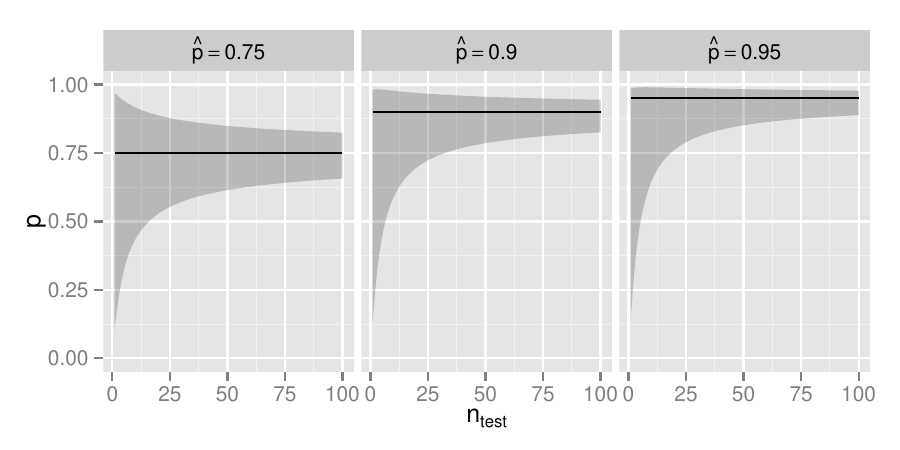}
\caption{95\,\% confidence intervals for different observed performances $\hat p$ as function of
  \ntest. If 90 out of 100 samples of a class are recognized correctly (\eg sensitivity of the
  leukocytes with 25 training samples), the 95\,\% confidence interval for the sensitivity ranges
  from %
  0.83 %
  to %
  0.94 %
  -- which in the context of our classification task reads as being between ``quite bad'' and
  ``really good''.}
  \label{fig:ci}
\end{figure}

Figure~\ref{fig:ci} shows the 95\,\% confidence intervals for different observed performances as
function of the test sample size. For our example application, \eg the sensitivity of the leukocyte
class reaches 0.90 rather quickly. If that model were tested with 100 leukocytes (\ie four times as
many as in our largest small data sets) and 90 of them were correctly recognized, the 95\,\%
confidence interval would range from %
0.83 %
(which would be considered quite bad as leukocytes are fairly easy to recognize) to %
0.94 %
-- which in the context of our classification task would be translated to ``quite good''. In other
words, the confidence interval would still be too wide to allow a practical judgment of the
classifier.

\begin{figure}[tb]
\includegraphics[width=\hlw]{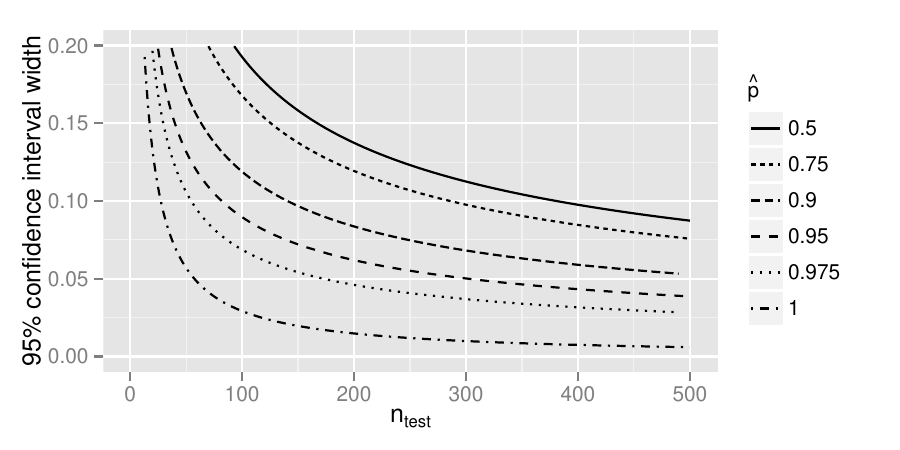}
  \caption{95\,\% confidence interval widths for different observed performances $\hat p$ as function
    of \ntest. $\hat p = 0.5$ and $1$ give the widest and narrowest possible confidence interval
    widths. \Eg If the confidence interval should not be more than 0.1 wide while a sensitivity of
    0.9 is expected, $\ntest \geq 141$
    samples need to be tested.}
  \label{fig:ciwidth}
\end{figure}

Similarly, already with 4 -- 5 training spectra (out of 6 total red blood cell spectra in the data
set), we observed perfect recognition of red blood cells in the simulation example's cross
validation. But the 95\,\% confidence interval still reaches down to %
0.65. %
However, for $\hat p = 1$ the confidence intervals narrow very soon, and ``already'' with %
58 %
test samples the lower limit of the 95\,\% confidence interval reaches 0.95 (see
fig.~\ref{fig:ciwidth}).

Figure~\ref{fig:ciwidth} gives the width of the Bayesian confidence interval as function of the test
sample size for different observed values of the performance. Note that specifying confidence
interval widths to be less than 0.10 with expected observed performance between 0.90 and 0.95 already
corresponds to requiring between 3 -- 5$\frac{1}{2}$ times as many test samples as we consider
typically available in biospectroscopy. For confidence interval widths of less than 0.05 which would
allow to distinguish the practical categories ``bad'' and ``very good'', hundreds of test cases are
required. Also, this estimation of required sample sizes is very sensitive to the true proportion $p$:
if $p$ were in fact only 0.89 instead of the 0.9 assumed in the example, %
153 %
instead of %
141 %
test samples would be required to reach the specified confidence interval width.

\subsection{Demonstrating that a New Classifier is Better}
\label{sec:superiority}

A second important scenario that allows to specify necessary test sample sizes is demonstrating
superiority to an already known classifier. \Eg, the instrument is improved and the resulting
advantage should be demonstrated. A rough estimate of the performance of the new instrument is
available. How many samples are needed in order to prove the superiority of the new approach?

From a statistics point of view, comparing classifier performance is a typical hypothesis testing
task.  R package Hmisc \cite{Hmisc} provides functions for power (\texttt{bpower}) and sample size
estimation (\texttt{bsamsize}) of independent proportions with unequal test sample sizes as described
by Fleiss \emph{et\,al.} \cite{Fleiss1980}. The approximation overestimates power for small sample sizes
\cite{Vorburger2006}. However, this is not of much consequence here, as the calculated sample sizes
will anyways be rough guesstimates rather than exact numbers of required samples: Firstly, the exact
performance of the improved classifier is unknown, so the sample size planning needs to check the
sensitivity of the calculated numbers to this assumption. Secondly, the actual power of the
calculated scenario can be checked by \texttt{bpower.sim}.

Assume our recognition of BT-20 cells were improved from the 0.75 sensitivity we obtain with 20
training samples\,/\,class to 0.90. A quick estimate of the necessary test sample size reveals that
in this scenario, the maximal obtainable power \footnote{Probability that we correctly conclude that
  the new classifier is better than the old one iff it actually is.} (setting \ntest for the new
model to 10\textsuperscript{5} as infinite for practical purposes) is $1 - \beta = $ %
0.62. %
In other words, there is no chance to prove the superiority of the new classifier with anything close
to an acceptable type II error\footnote{Probability that we wrongly conclude the new classifier is no
  better than the old one, although it actually is.} due to the small test sample size available for
the old model.  The comparisons have most power if the tests are performed with equal sample
sizes. For this case, tables are also available in Fleiss and Paik
\cite{StatisticalMethodsForRatesAndProportions-Fleiss}. In our example, the usual power of 0.8 (\ie
type II error $\beta = 0.2$; with type I error\footnote{Probability that we wrongly conclude the new
  classifier is better than the old one, although it actually is not.}$ \alpha = 0.05$) needs at
least %
100 %
independent test cases truly belonging to class bt for each of the models. Note that paired tests can
be much more powerful, thus requiring less samples. Paired tests can be used when the same cases can
be measured again (impossible for our study: new cell culture batches need to be grown) or if the
improvement is in the data analysis and therefore the same instrumental data can be analysed by both
methods.

\begin{figure}[t]
\includegraphics[width=\hlw]{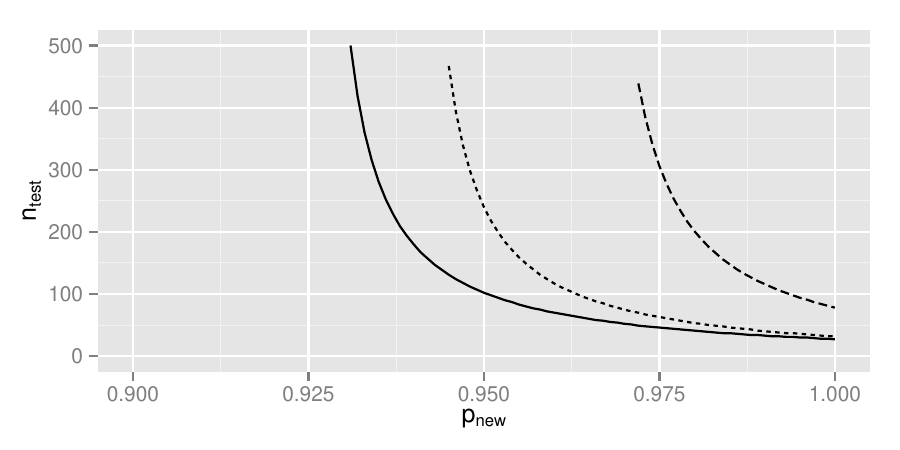}
\caption{\label{fig:improve} Test sample size necessary to demonstrate superiority of an improved
  model of sensitivity $p_{new}$, assuming the ``old'' model had $p_{old}$ = 0.75 sensitivity and
  was tested with \ntest = 25 samples and accepting a type I error of $\alpha = 0.05$ and a type II
  error $\beta = 0.2$ (solid line).  Dotted: test sample size for the second model with $\alpha =
  \beta = 0.10$.  However, if $\alpha = \beta = 0.05$ is required (dashed), even a model with
  0.975 true sensitivity needs to be tested with at least 300 cases and 116 cases are necessary to
  demonstrate the superiority of an improved model truly achieving 0.99 sensitivity.}
\end{figure}
If we could achieve 0.975 sensitivity for BT-20 cells, we would need to test with 
63 %
test cases (accepting $\alpha = \beta = 0.10$). Figure~\ref{fig:improve} shows that this is very
sensitive to the assumed quality of the new model: if the new model has in fact ``only'' a sensitivity of 0.96 
 (corresponding to ca. 1 additional misclassification out of 63 test cases), already %
117 %
or almost twice as many test cases are needed.  Note that this is a rather extreme example as it
means one order of magnitude (0.25 to 0.025) reduction in the fraction of unrecognized BT-20
cells, which is much larger than the improvements considered in the practice of biospectroscopic
classification.

In conclusion, well working classifiers need to be validated with at least 75 test cases in order to
obtain confidence intervals that are narrow enough to draw practical conclusions about the 
model. Demonstrating superiority of a new, improved classifier in general needs even more test cases
and often will be impossible at all if the test sample size for the old classifier was small.
\section{Summary}
\label{sec:conclusions}
Using a Raman spectroscopic five class classification problem as well as simulated data based on the
real data set, we compared the sample sizes needed to train good classifiers with sample sizes needed
to demonstrate that the obtained classifiers work well. Due to the smaller test sample size,
sensitivities are more difficult to determine precisely than specificities or overall hit rates.

Using typical small sample sizes of up to 25 samples per class, we calculated learning curves
(sensitivity as function of the training sample size) using 100$\times$ iterated $5$-fold cross
validation. While the general shape of the learning curve could be determined correctly for the very
easily recognized red blood cells, for more difficult recognition tasks not even the correct shape of
the learning curve can be determined reliably within the small data set as the precise measurement of
classifier performance requires rather large test sample sizes (> 75 cases).

In consequence, we calculate necessary test sample sizes for different pre-specified testing
scenarios, namely specifying acceptable widths for the confidence interval of the true sensitivity
and the number of test samples needed to demonstrate superiority of one classifier over another. In
order to obtain confidence interval widths $\leq$ 0.1, 140 test samples are necessary when 90\,\%
sensitivity is expected. In contrast, the recognition of leukocytes in our example application
reaches 90\,\% sensitivity already with about 20 training samples. Comparison of classifiers was
found to require even larger test sample sizes (hundreds of statistically independent cases) in the
general case.

In conclusion, we recommend to start sample size planning for classification by specifying acceptable
confidence interval widths for the expected sensitivities. This will lead to sample sizes that allow
retrospective calculation of learning curves and a refined sample size planning in terms of both test
and training sample size can then be done.

\section*{Acknowledgments}
Graphics were generated using ggplot2 \cite{ggplot2}.

Financial support by the European Union via the Europäischer Fonds für Regionale Entwicklung (EFRE)
and the Thüringer Ministerium für Bildung, Wissenschaft und Kultur (project B714-07037) as well as
the funding by BMBF (FKZ 01EO1002) is highly acknowledged.

%% Local Variables: 
%% mode: ess-noweb
%% TeX-master: t
%% End: 

%% file: supplementary-confmat.tex
% This file constitutes the supplementary material to 
% 
% C. Beleites, U. Neugebauer, T. Bocklitz, C. Krafft, J. Popp: 
% Sample Size Planning for Classification Models
% submitted to: Analytica Chimica Acta, Special Issue CAC2012
%
% Version History & change log: 
% 2012-08-10 Original version
%     Claudia Beleites, IPHT Jena e.V., Albert-Einstein-Str. 9, D-07745 Jena, Germany 
% 
\section*{Supplementary file I: Confusion matrices for table 1}
We give here the complete confusion tables for the models described in table 1.
The counts are divided by the number of test samples truly belonging to the respective class and averaged over all runs, i.\,e. the diagonal elements are the sensitivities, the other entries are the specificities with respect to that particular misclassification.

\subsection*{Bayes Performance for Simulated Data LDA Model}

\begin{knitrout}
\definecolor{shadecolor}{rgb}{0.969, 0.969, 0.969}\color{fgcolor}\begin{kframe}
\begin{alltt}
conf.mat.bayes
\end{alltt}
\begin{verbatim}
##      pred
## ref   rbc leu  mcf   bt  oci
##   rbc   1   0 0.00 0.00 0.00
##   leu   0   1 0.00 0.00 0.00
##   mcf   0   0 0.95 0.03 0.02
##   bt    0   0 0.04 0.91 0.05
##   oci   0   0 0.02 0.03 0.94
\end{verbatim}
\begin{alltt}
n
\end{alltt}
\begin{verbatim}
##   rbc   leu   mcf    bt   oci 
## 40000 40000 40000 40000 40000
\end{verbatim}
\end{kframe}
\end{knitrout}

\subsection*{Bayes Performance for Simulated Data 10 latent variable PLS-LDA}

\begin{knitrout}
\definecolor{shadecolor}{rgb}{0.969, 0.969, 0.969}\color{fgcolor}\begin{kframe}
\begin{alltt}
conf.mat.plsbayes
\end{alltt}
\begin{verbatim}
##      pred
## ref   rbc  leu  mcf   bt  oci
##   rbc   1 0.00 0.00 0.00 0.00
##   leu   0 0.99 0.01 0.00 0.01
##   mcf   0 0.00 0.87 0.06 0.07
##   bt    0 0.00 0.14 0.72 0.13
##   oci   0 0.00 0.04 0.09 0.86
\end{verbatim}
\begin{alltt}
n
\end{alltt}
\begin{verbatim}
##   rbc   leu   mcf    bt   oci 
## 40000 40000 40000 40000 40000
\end{verbatim}
\end{kframe}
\end{knitrout}

\subsection*{Average ``Bayes'' Performance for Real Data as used in the paper}

\begin{knitrout}
\definecolor{shadecolor}{rgb}{0.969, 0.969, 0.969}\color{fgcolor}\begin{kframe}
\begin{alltt}
confmat.bayes
\end{alltt}
\begin{verbatim}
##      pred
## ref    rbc  leu  mcf   bt  oci
##   rbc 0.99 0.01 0.00 0.00 0.00
##   leu 0.00 0.97 0.01 0.00 0.02
##   mcf 0.00 0.00 0.91 0.04 0.05
##   bt  0.00 0.00 0.13 0.75 0.13
##   oci 0.00 0.01 0.04 0.07 0.89
\end{verbatim}
\begin{alltt}
n
\end{alltt}
\begin{verbatim}
## rbc leu mcf  bt oci 
## 372 569 558 532 518
\end{verbatim}
\end{kframe}
\end{knitrout}

\subsection*{Average ``Bayes'' Performance for Real Data batch-wise validation}

\begin{knitrout}
\definecolor{shadecolor}{rgb}{0.969, 0.969, 0.969}\color{fgcolor}\begin{kframe}
\begin{alltt}
confmat.batch
\end{alltt}
\begin{verbatim}
##      pred
## ref    rbc  leu  mcf   bt  oci
##   rbc 0.98 0.02 0.00 0.00 0.00
##   leu 0.00 0.87 0.03 0.04 0.06
##   mcf 0.00 0.00 0.33 0.37 0.30
##   bt  0.00 0.00 0.32 0.38 0.30
##   oci 0.00 0.02 0.31 0.36 0.31
\end{verbatim}
\begin{alltt}
n
\end{alltt}
\begin{verbatim}
## rbc leu mcf  bt oci 
## 372 569 558 532 518
\end{verbatim}
\end{kframe}
\end{knitrout}

This file is supplementary material for: \href{http://dx.doi.org/10.1016/j.aca.2012.11.007}{C.~Beleites, U.~Neugebauer, T.~Bocklitz, C.~Krafft and J.~Popp: \emph{Sample size planning for classification models}. Analytica Chimica Acta, 2013, 760 (Special Issue: Chemometrics in Analytical Chemistry 2012), 25--33, DOI: 10.1016/j.aca.2012.11.007}.

%% file: supplementary-code.tex
% This file constitutes the supplementary material to 
% 
% C. Beleites, U. Neugebauer, T. Bocklitz, C. Krafft, J. Popp: 
% Sample Size Planning for Classification Models
% submitted to: Analytica Chimica Acta, Special Issue CAC2012
%
% License: Creative Commons Attribution-ShareAlike 3.0 Unported (CC BY-SA 3.0) 
%          http://creativecommons.org/licenses/by-sa/3.0/
% 
% Version History & change log: 
% 2012-08-08 Original version 
%     Claudia Beleites, IPHT Jena e.V., Albert-Einstein-Str. 9, D-07745 Jena, Germany 
% 2012-08-10 added license 
%     Claudia Beleites, IPHT Jena e.V., Albert-Einstein-Str. 9, D-07745 Jena, Germany 
% 2013-01-04 make file appendix for arXiv version of the manuscript
%     Claudia Beleites, IPHT Jena e.V., Albert-Einstein-Str. 9, D-07745 Jena, Germany 
%
% BEGIN UNCOMMENT FOR STANDALONE DOCUMENT ----------------------------------------------------------
% AND REMOVE SPACE IN DOCUMENTCLASS
%  |
%  v
% \ documentclass[10pt, A4paper, headings=small, DIV14]{scrartcl}
% \usepackage{inputenx}
% \usepackage{url}
% \usepackage{xspace}
% \usepackage{ccicons}
% \newcommand{\ntest}{\ensuremath{n_{test}}\xspace}
% \newcommand{\eg}{\emph{e.\,g.}\xspace}
% \newcommand{\Eg}{\emph{E.\,g.}\xspace}
% \begin{document}
% \appendix
%
% END UNCOMMENT FOR STANDALONE DOCUMENT ----------------------------------------------------------
\section*{Supplementary file II: \texttt{R} code for section 4}
\subsection*{Specifying Acceptable Confidence Interval Widths}

\subsubsection*{Calculation of the Bayesian confidence intervals and their widths}
\label{sec:calc-bayes-conf}
\begin{knitrout}
\definecolor{shadecolor}{rgb}{0.969, 0.969, 0.969}\color{fgcolor}\begin{kframe}
\begin{alltt}
\hlfunctioncall{require}(\hlstring{"binom"})
\hlfunctioncall{require}(\hlstring{"ggplot2"})
\end{alltt}
\end{kframe}
\end{knitrout}

We consider test sample sizes up to 500:
\begin{knitrout}
\definecolor{shadecolor}{rgb}{0.969, 0.969, 0.969}\color{fgcolor}\begin{kframe}
\begin{alltt}
n <- 1:500
\end{alltt}
\end{kframe}
\end{knitrout}

Now calculate confidence intervals for different observed performance values $\hat p$ between 0.5 and
1. As \texttt{binom.bayes} does not converge for all combinations of $n$ and $\hat p$, keep only
results where the actual confidence level is 0.95 $\pm 10^{-4}$ (a bunch of warnings will be
thrown; filter the results without adequate coverage afterwards). The uniform prior is selected by setting both \texttt{prior.shape1} and \texttt{prior.shape2} to 1.
\begin{knitrout}
\definecolor{shadecolor}{rgb}{0.969, 0.969, 0.969}\color{fgcolor}\begin{kframe}
\begin{alltt}
confint <- \hlfunctioncall{lapply}(\hlfunctioncall{c}(0.5, 0.75, 0.9, 0.95, 0.975, 1), \hlfunctioncall{function}(p) \{
    tmp <- \hlfunctioncall{binom.bayes}(n = n, x = p * n, prior.shape1 = 1, prior.shape2 = 1)
    tmp$phat <- p
    \hlfunctioncall{subset}(tmp, sig >= 0.05 - 1e-04 & sig <= 0.05 + 1e-04)
\})
confint <- \hlfunctioncall{do.call}(rbind, confint)
\end{alltt}
\end{kframe}
\end{knitrout}

Add column containing the width of the confidence interval:
\begin{knitrout}
\definecolor{shadecolor}{rgb}{0.969, 0.969, 0.969}\color{fgcolor}\begin{kframe}
\begin{alltt}
confint$width <- confint$upper - confint$lower
\end{alltt}
\end{kframe}
\end{knitrout}

\begin{knitrout}
\definecolor{shadecolor}{rgb}{0.969, 0.969, 0.969}\color{fgcolor}\begin{kframe}
\begin{alltt}
\hlfunctioncall{head}(confint)
\end{alltt}
\begin{verbatim}
##   method   x n shape1 shape2 mean   lower  upper  sig phat  width
## 1  bayes 0.5 1    1.5    1.5  0.5 0.06083 0.9392 0.05  0.5 0.8783
## 2  bayes 1.0 2    2.0    2.0  0.5 0.09430 0.9057 0.05  0.5 0.8114
## 3  bayes 1.5 3    2.5    2.5  0.5 0.12275 0.8772 0.05  0.5 0.7545
## 4  bayes 2.0 4    3.0    3.0  0.5 0.14663 0.8534 0.05  0.5 0.7067
## 5  bayes 2.5 5    3.5    3.5  0.5 0.16681 0.8332 0.05  0.5 0.6664
## 6  bayes 3.0 6    4.0    4.0  0.5 0.18405 0.8159 0.05  0.5 0.6319
\end{verbatim}
\end{kframe}
\end{knitrout}

Sample sizes used in the text:
\begin{knitrout}
\definecolor{shadecolor}{rgb}{0.969, 0.969, 0.969}\color{fgcolor}\begin{kframe}
\begin{alltt}
(nleuko <- \hlfunctioncall{min}(\hlfunctioncall{subset}(confint, phat == 0.9 & width <= 0.1)$n))
\end{alltt}
\begin{verbatim}
## [1] 141
\end{verbatim}
\begin{alltt}
(nery <- \hlfunctioncall{min}(\hlfunctioncall{subset}(confint, phat == 1 & width <= 0.05)$n))
\end{alltt}
\begin{verbatim}
## [1] 58
\end{verbatim}
\begin{alltt}

tmp <- \hlfunctioncall{binom.bayes}(n = n, x = 0.89 * n, prior.shape1 = 1, prior.shape2 = 1)
tmp$width <- tmp$upper - tmp$lower
(nleuko89 <- \hlfunctioncall{min}(\hlfunctioncall{subset}(tmp, width <= 0.1)$n))
\end{alltt}
\begin{verbatim}
## [1] 153
\end{verbatim}
\end{kframe}
\end{knitrout}

95\,\% confidence interval for 90 correct out of 100 tested leukocytes:   
\begin{knitrout}
\definecolor{shadecolor}{rgb}{0.969, 0.969, 0.969}\color{fgcolor}\begin{kframe}
\begin{alltt}
\hlfunctioncall{binom.bayes}(n = 100, x = 90, prior.shape1 = 1, prior.shape2 = 1)
\end{alltt}
\begin{verbatim}
##   method  x   n shape1 shape2   mean  lower  upper  sig
## 1  bayes 90 100     91     11 0.8922 0.8254 0.9444 0.05
\end{verbatim}
\end{kframe}
\end{knitrout}

95\,\% confidence interval for 6 correct out of 6 tested erythrocytes:   
\begin{knitrout}
\definecolor{shadecolor}{rgb}{0.969, 0.969, 0.969}\color{fgcolor}\begin{kframe}
\begin{alltt}
\hlfunctioncall{binom.bayes}(n = 6, x = 6, prior.shape1 = 1, prior.shape2 = 1)
\end{alltt}
\begin{verbatim}
##   method x n shape1 shape2  mean  lower upper  sig
## 1  bayes 6 6      7      1 0.875 0.6518     1 0.05
\end{verbatim}
\end{kframe}
\end{knitrout}

Generate the plots:
\begin{knitrout}
\definecolor{shadecolor}{rgb}{0.969, 0.969, 0.969}\color{fgcolor}\begin{kframe}
\begin{alltt}
\hlfunctioncall{ggplot} (\hlfunctioncall{subset} (confint, phat %in% \hlfunctioncall{c} (.75, .9, .95) & n <= 100)) + 
  \hlfunctioncall{geom_ribbon} (\hlfunctioncall{aes} (x = n, ymin = lower, ymax = upper), alpha = 0.25) + 
  \hlfunctioncall{facet_grid} (. ~ phat, labeller = \hlfunctioncall{label_bquote}(expr = \hlfunctioncall{hat} (p) == \hlfunctioncall{.}(x))) + 
  \hlfunctioncall{geom_line} (\hlfunctioncall{aes} (x = n, y = phat)) +
  \hlfunctioncall{ylab} (\hlstring{"p"}) + \hlfunctioncall{xlab} (\hlfunctioncall{expression} (n [test])) 
\end{alltt}
\end{kframe}\includegraphics[width=\maxwidth]{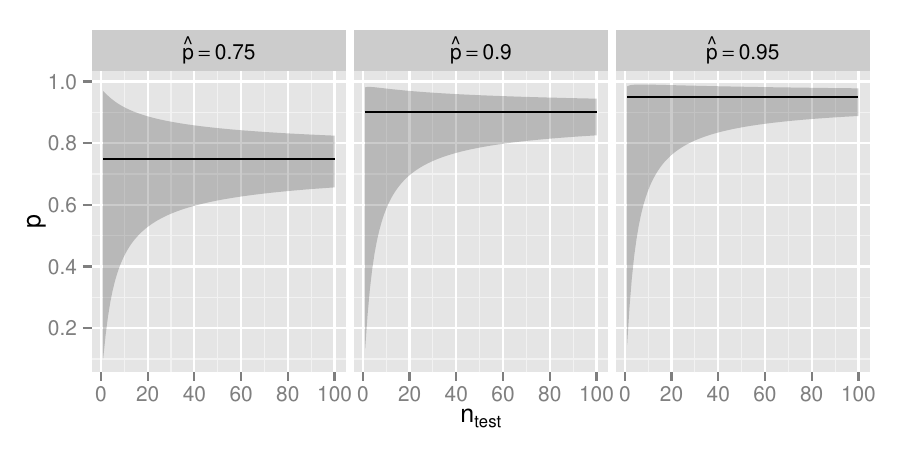} 
\end{knitrout}

\begin{knitrout}
\definecolor{shadecolor}{rgb}{0.969, 0.969, 0.969}\color{fgcolor}\begin{kframe}
\begin{alltt}
\hlfunctioncall{ggplot} (\hlfunctioncall{subset} (confint, width <= 0.2)) + 
  \hlfunctioncall{geom_line} (\hlfunctioncall{aes} (x = n, y = width, lty = \hlfunctioncall{as.factor} (phat))) +
  \hlfunctioncall{scale_linetype_discrete} (\hlfunctioncall{expression} (\hlfunctioncall{hat} (p))) +
  \hlfunctioncall{ylim} (0, .2) + \hlfunctioncall{ylab} (\hlstring{"95% confidence interval width"}) + 
  \hlfunctioncall{xlab} (\hlfunctioncall{expression} (n [test])) 
\end{alltt}
\end{kframe}\includegraphics[width=\maxwidth]{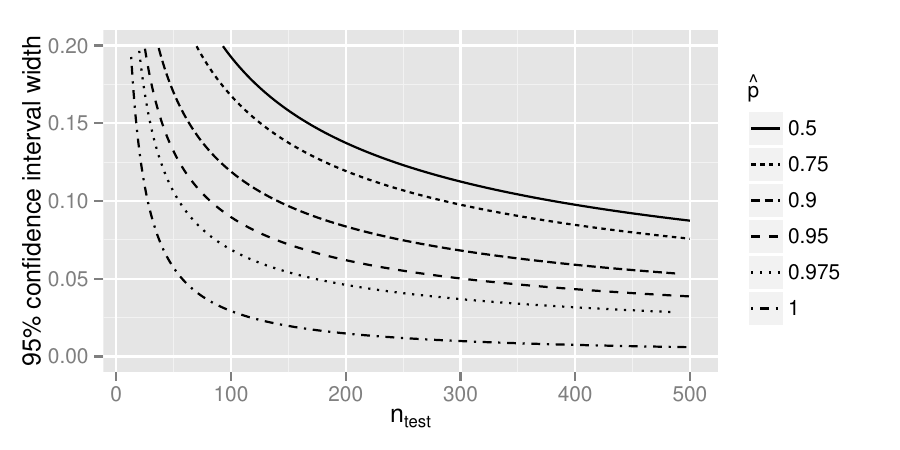} 
\end{knitrout}

\subsection*{Demonstrating that a New Classifier is Better}
\begin{knitrout}
\definecolor{shadecolor}{rgb}{0.969, 0.969, 0.969}\color{fgcolor}\begin{kframe}
\begin{alltt}
\hlfunctioncall{require}(\hlstring{"Hmisc"})
\end{alltt}
\end{kframe}
\end{knitrout}

Assume our recognition of BT-20 cells were improved from the 75\,\% sensitivity we obtain with 20
training samples\,/\,class to 90\,\% (Bayes error for LDA of the simulated data). A quick estimate of
the necessary test sample size reveals that in this scenario, the maximal obtainable
power (setting \ntest for the new model to 10\textsuperscript{5} as infinite for practical
purposes) is $1 - \beta = $ %
\begin{knitrout}
\definecolor{shadecolor}{rgb}{0.969, 0.969, 0.969}\color{fgcolor}\begin{kframe}
\begin{alltt}
\hlfunctioncall{bpower.sim}(p1 = 0.75, p2 = 0.9, n1 = 25, n2 = 1e+05, nsim = 1e+05)
\end{alltt}
\begin{verbatim}
##  Power  Lower  Upper 
## 0.6218 0.6188 0.6248
\end{verbatim}
\end{kframe}
\end{knitrout}

Required sample size for equal test sample sizes for both classifiers:
\begin{knitrout}
\definecolor{shadecolor}{rgb}{0.969, 0.969, 0.969}\color{fgcolor}\begin{kframe}
\begin{alltt}
\hlfunctioncall{bsamsize}(p1 = 0.75, p2 = 0.9)
\end{alltt}
\begin{verbatim}
##    n1    n2 
## 99.54 99.54
\end{verbatim}
\end{kframe}
\end{knitrout}

Prepare necessary sample size if old classifier was tested with 25 samples. \texttt{bsamsize} only
accepts fractions of samples, not directly specifying $n_1$ = $n_{old}$. Thus, we need to find the
fraction that corresponds to $n_{old}$ = 25 and from there calculate $n_{new}$.
\begin{knitrout}
\definecolor{shadecolor}{rgb}{0.969, 0.969, 0.969}\color{fgcolor}\begin{kframe}
\begin{alltt}
target.fun <- \hlfunctioncall{function}(fraction, ..., n.old) \{
    n1 <- \hlfunctioncall{bsamsize}(fraction = fraction, ...)[1]
    (n1 - n.old)^2
\}

est.n2 <- \hlfunctioncall{function}(p2, ..., n.old) \{
    fraction <- \hlfunctioncall{optimize}(target.fun, lower = 1e-05, upper = 0.5, ..., p2 = p2, 
        n.old = n.old)$minimum
    \hlfunctioncall{ceiling}(n.old/fraction - n.old)
\}
\end{alltt}
\end{kframe}
\end{knitrout}

We consider $p_1 = 0.75, p_2 \in [0.9, 1]$ and $n_{old} = 25$ (target value for $n_1$) for the three
different type I and type II errors discussed in the text:
\begin{knitrout}
\definecolor{shadecolor}{rgb}{0.969, 0.969, 0.969}\color{fgcolor}\begin{kframe}
\begin{alltt}
p2 <- \hlfunctioncall{seq}(0.9, 1, by = 0.001)

n2 <- \hlfunctioncall{sapply}(p2, est.n2, power = 0.95, p1 = 0.75, n.old = 25)
df <- \hlfunctioncall{data.frame}(p2 = p2, n2 = n2, power = 0.95)

n2 <- \hlfunctioncall{sapply}(p2, est.n2, power = 0.9, alpha = 0.1, p1 = 0.75, n.old = 25)
df <- \hlfunctioncall{rbind}(df, \hlfunctioncall{data.frame}(p2 = p2, n2 = n2, power = 0.9))

n2 <- \hlfunctioncall{sapply}(p2, est.n2, power = 0.8, p1 = 0.75, n.old = 25)
df <- \hlfunctioncall{rbind}(df, \hlfunctioncall{data.frame}(p2 = p2, n2 = n2, power = 0.8))
\end{alltt}
\end{kframe}
\end{knitrout}

\begin{knitrout}
\definecolor{shadecolor}{rgb}{0.969, 0.969, 0.969}\color{fgcolor}\begin{kframe}
\begin{alltt}
\hlfunctioncall{ggplot} (data = \hlfunctioncall{subset} (df, n2 <= 500), \hlfunctioncall{aes} (x = p2, y = n2, lty = \hlfunctioncall{as.factor} (power))) + 
  \hlfunctioncall{geom_line} () + \hlfunctioncall{ylim} (0, 500) + \hlfunctioncall{xlim} (0.90, 1) +
  \hlfunctioncall{xlab} (\hlfunctioncall{expression} (p [new])) + \hlfunctioncall{ylab} (\hlfunctioncall{expression} (n[test,~new])) 
\end{alltt}
\end{kframe}\includegraphics[width=\maxwidth]{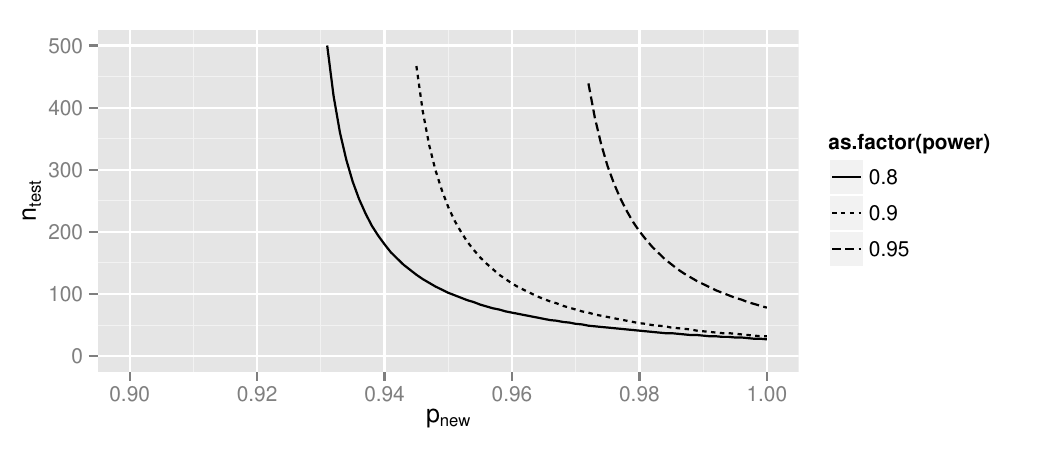} 
\end{knitrout}

The example values are extracted from the data.frame in the paper:

\begin{knitrout}
\definecolor{shadecolor}{rgb}{0.969, 0.969, 0.969}\color{fgcolor}\begin{kframe}
\begin{alltt}
\hlfunctioncall{subset}(df, p2 == 0.975 & power == 0.9, n2)
\end{alltt}
\begin{verbatim}
##     n2
## 177 63
\end{verbatim}
\begin{alltt}
\hlfunctioncall{subset}(df, p2 == 0.96 & power == 0.9, n2)
\end{alltt}
\begin{verbatim}
##      n2
## 162 117
\end{verbatim}
\end{kframe}
\end{knitrout}

but can also be calculated directly with the functions defined above:
\begin{knitrout}
\definecolor{shadecolor}{rgb}{0.969, 0.969, 0.969}\color{fgcolor}\begin{kframe}
\begin{alltt}
\hlfunctioncall{est.n2}(p1 = 0.75, p2 = 0.975, power = 0.9, alpha = 0.1, n.old = 25)
\end{alltt}
\begin{verbatim}
## [1] 63
\end{verbatim}
\begin{alltt}
\hlfunctioncall{est.n2}(p1 = 0.75, p2 = 0.96, power = 0.9, alpha = 0.1, n.old = 25)
\end{alltt}
\begin{verbatim}
## [1] 117
\end{verbatim}
\end{kframe}
\end{knitrout}

\subsection*{Information on Using and Compiling this file}
\label{sec:information}

This document is provided in two forms. The \texttt{.Rnw} is a \texttt{knitr} (Sweave) file. It can
be processed in R using:
\begin{knitrout}
\definecolor{shadecolor}{rgb}{0.969, 0.969, 0.969}\color{fgcolor}\begin{kframe}
\begin{alltt}
\hlfunctioncall{require}(\hlstring{"knitr"})
\hlfunctioncall{knit}(\hlstring{"supplementary-code.Rnw"})
\end{alltt}
\end{kframe}
\end{knitrout}

which produces a \LaTeX\ document that can be compiled into the \texttt{.pdf} by \texttt{pdflatex} (uncomment the document header and footer as indicated). 

\subsubsection*{License}
You may use this document under the Creative Commons CC BY-SA License version 3.0 (see \url{http://creativecommons.org/licenses/by-sa/3.0/})

{\centering{\Huge\ccbysa}\\}

As attribution, please cite: \href{http://dx.doi.org/10.1016/j.aca.2012.11.007}{C.~Beleites, U.~Neugebauer, T.~Bocklitz, C.~Krafft and J.~Popp: \emph{Sample size planning for classification models}. Analytica Chimica Acta, 2013, 760 (Special Issue: Chemometrics in Analytical Chemistry 2012), 25--33, DOI: 10.1016/j.aca.2012.11.007}.

\subsubsection*{\texttt{R} Session Information}
\begin{knitrout}
\definecolor{shadecolor}{rgb}{0.969, 0.969, 0.969}\color{fgcolor}\begin{kframe}
\begin{alltt}
\hlfunctioncall{sessionInfo}()
\end{alltt}
\begin{verbatim}
## R version 2.15.2 (2012-10-26)
## Platform: x86_64-pc-linux-gnu (64-bit)
## 
## locale:
##  [1] LC_CTYPE=de_DE.UTF-8       LC_NUMERIC=C              
##  [3] LC_TIME=de_DE.UTF-8        LC_COLLATE=de_DE.UTF-8    
##  [5] LC_MONETARY=de_DE.UTF-8    LC_MESSAGES=de_DE.UTF-8   
##  [7] LC_PAPER=C                 LC_NAME=C                 
##  [9] LC_ADDRESS=C               LC_TELEPHONE=C            
## [11] LC_MEASUREMENT=de_DE.UTF-8 LC_IDENTIFICATION=C       
## 
## attached base packages:
## [1] splines   stats     graphics  grDevices utils     datasets  methods  
## [8] base     
## 
## other attached packages:
## [1] Hmisc_3.10-1     survival_2.36-14 ggplot2_0.9.2.1  binom_1.0-5     
## [5] lattice_0.20-10  knitr_0.8       
## 
## loaded via a namespace (and not attached):
##  [1] cluster_1.14.3     colorspace_1.2-0   dichromat_1.2-4   
##  [4] digest_0.5.2       evaluate_0.4.2     formatR_0.6       
##  [7] grid_2.15.2        gtable_0.1.2       labeling_0.1      
## [10] MASS_7.3-22        memoise_0.1        munsell_0.4       
## [13] plyr_1.7.1         proto_0.3-9.2      RColorBrewer_1.0-5
## [16] reshape2_1.2.1     scales_0.2.1       stringr_0.6.1     
## [19] tools_2.15.2
\end{verbatim}
\end{kframe}
\end{knitrout}

% BEGIN UNCOMMENT FOR STANDALONE DOCUMENT ----------------------------------------------------------
%\end{document}
% END UNCOMMENT FOR STANDALONE DOCUMENT ------------------------------------------------------------

%% file: samplesize.bbl
\begin{thebibliography}{10}
\expandafter\ifx\csname url\endcsname\relax
  \def\url#1{\texttt{#1}}\fi
\expandafter\ifx\csname urlprefix\endcsname\relax\def\urlprefix{URL }\fi
\expandafter\ifx\csname href\endcsname\relax
  \def\href#1#2{#2} \def\path#1{#1}\fi

\bibitem{Mukherjee2003}
S.~Mukherjee, P.~Tamayo, S.~Rogers, R.~Rifkin, A.~Engle, C.~Campbell, T.~R.
  Golub, J.~P. Mesirov,
  \href{http://dx.doi.org/10.1089/106652703321825928}{Estimating dataset size
  requirements for classifying DNA microarray data.}, J Comput Biol 10~(2)
  (2003) 119--142.
\newblock \href {http://dx.doi.org/10.1089/106652703321825928}
  {\path{doi:10.1089/106652703321825928}}.

\bibitem{Figueroa2012}
R.~L. Figueroa, Q.~Zeng-Treitler, S.~Kandula, L.~H. Ngo,
  \href{http://dx.doi.org/10.1186/1472-6947-12-8}{Predicting sample size
  required for classification performance.}, BMC Med Inform Decis Mak 12~(1)
  (2012) 8.
\newblock \href {http://dx.doi.org/10.1186/1472-6947-12-8}
  {\path{doi:10.1186/1472-6947-12-8}}.

\bibitem{Dobbin2007}
K.~K. Dobbin, R.~M. Simon,
  \href{http://dx.doi.org/10.1093/biostatistics/kxj036}{Sample size planning
  for developing classifiers using high-dimensional DNA microarray data.},
  Biostatistics 8~(1) (2007) 101--117.
\newblock \href {http://dx.doi.org/10.1093/biostatistics/kxj036}
  {\path{doi:10.1093/biostatistics/kxj036}}.

\bibitem{Dobbin2008}
K.~K. Dobbin, Y.~Zhao, R.~M. Simon,
  \href{http://dx.doi.org/10.1158/1078-0432.CCR-07-0443}{How large a training
  set is needed to develop a classifier for microarray data?}, Clin Cancer Res
  14~(1) (2008) 108--114.
\newblock \href {http://dx.doi.org/10.1158/1078-0432.CCR-07-0443}
  {\path{doi:10.1158/1078-0432.CCR-07-0443}}.

\bibitem{Jain1982}
A.~Jain, B.~Chandrasekaran, Dimensionality and Sample Size Considerations in
  Pattern Recognition Practice, in: P.~R. Krishnaiah, L.~Kanal (Eds.), Handbook
  of Statistics, Vol.~II of Handbook of Statistics, North-Holland, Amsterdam,
  1982, Ch.~39, pp. 835 -- 855.

\bibitem{Raudys1991}
S.~Raudys, A.~Jain, Small Sample Size Effects in Statistical Pattern
  Recognition: Recommendations for Practitioners, IEEE Transactions on Pattern
  Analysis and Machine Intelligence 13 (1991) 252--264.
\newblock \href
  {http://dx.doi.org/http://doi.ieeecomputersociety.org/10.1109/34.75512}
  {\path{doi:http://doi.ieeecomputersociety.org/10.1109/34.75512}}.

\bibitem{Kalayeh1983}
H.~M. Kalayeh, D.~A. Landgrebe, Predicting the required number of training
  samples., IEEE Trans Pattern Anal Mach Intell 5~(6) (1983) 664--667.

\bibitem{Neugebauer2010}
U.~Neugebauer, T.~Bocklitz, J.~H. Clement, C.~Krafft, J.~Popp,
  \href{http://dx.doi.org/10.1039/c0an00608d}{Towards detection and
  identification of circulating tumour cells using Raman spectroscopy.},
  Analyst 135~(12) (2010) 3178--3182.
\newblock \href {http://dx.doi.org/10.1039/c0an00608d}
  {\path{doi:10.1039/c0an00608d}}.

\bibitem{Neugebauer2010a}
U.~Neugebauer, J.~H. Clement, T.~Bocklitz, C.~Krafft, J.~Popp,
  \href{http://dx.doi.org/10.1002/jbio.201000020}{Identification and
  differentiation of single cells from peripheral blood by Raman spectroscopic
  imaging.}, J Biophotonics 3~(8-9) (2010) 579--587.
\newblock \href {http://dx.doi.org/10.1002/jbio.201000020}
  {\path{doi:10.1002/jbio.201000020}}.

\bibitem{Hastie2009}
T.~Hastie, R.~Tibshirani, J.~Friedman, The Elements of Statistical Learning;
  Data mining, Inference andPrediction, 2nd Edition, Springer Verlag, New York,
  2009.

\bibitem{Dougherty2010}
E.~R. Dougherty, C.~Sima, J.~Hua, B.~Hanczar, U.~M. Braga-Neto,
  \href{http://www.ece.tamu.edu/~ulisses/public/Dougherty_CB_2010_preprint.pdf}{Performance
  of Error Estimators for Classification}, Current Bioinformatics 5 (2010)
  53--67.

\bibitem{Kohavi1995}
R.~Kohavi, A Study of Cross-Validation and Bootstrap for Accuracy Estimation
  and Model Selection, in: C.~S. Mellish (Ed.), Artificial Intelligence
  Proceedings 14$^{th}$ International Joint Conference, 20 -- 25. August 1995,
  Montr{\'e}al, Qu{\'e}bec, Canada, Morgan Kaufmann, USA, 1995, pp. 1137 --
  1145.

\bibitem{Beleites2005}
C.~Beleites, R.~Baumgartner, C.~Bowman, R.~Somorjai, G.~Steiner, R.~Salzer,
  M.~G. Sowa, Variance reduction in estimating classification error using
  sparse datasets, Chem.Intell.Lab.Syst. 79 (2005) 91 -- 100.

\bibitem{Esbensen2010}
K.~H. Esbensen, P.~Geladi, \href{http://dx.doi.org/10.1002/cem.1310}{Principles
  of Proper Validation: use and abuse of re-sampling for validation}, J.
  Chemometrics 24~(3-4) (2010) 168--187.

\bibitem{R}
{R Development Core Team}, \href{http://www.R-project.org/}{R: A Language and
  Environment for Statistical Computing}, R Foundation for Statistical
  Computing, Vienna, Austria, {ISBN} 3-900051-07-0 (2011).

\bibitem{hyperSpec}
C.~Beleites, V.~Sergo,
  \href{http://hyperspec.r-forge.r-project.org}{{hyperSpec}: a package to
  handle hyperspectral data sets in R}, {R} package v. 0.98-20120725 (2012).

\bibitem{mvtnorm.1}
A.~Genz, F.~Bretz, T.~Miwa, X.~Mi, F.~Leisch, F.~Scheipl, T.~Hothorn,
  \href{http://CRAN.R-project.org/package=mvtnorm}{{mvtnorm}: Multivariate
  Normal and t Distributions}, {R} package v. 0.9-9992 (2012).

\bibitem{mvtnorm.2}
A.~Genz, F.~Bretz, Computation of Multivariate Normal and t Probabilities,
  Lecture Notes in Statistics, Springer-Verlag, Heidelberg, 2009.

\bibitem{cbmodels}
C.~Beleites, cbmodels: Collection of "combined" models: PCA-LDA, PLS-LDA, etc.,
  {R} package v. 0.5-20120731 (2012).

\bibitem{pls}
R.~Wehrens, B.-H. Mevik, \href{http://mevik.net/work/software/pls.html}{pls:
  Partial Least Squares Regression (PLSR) and Principal Component Regression
  (PCR)}, R package version 2.1-0 (2007).

\bibitem{MASS}
W.~N. Venables, B.~D. Ripley, \href{http://www.stats.ox.ac.uk/pub/MASS4}{Modern
  Applied Statistics with S}, 4th Edition, Springer, New York, 2002, ISBN
  0-387-95457-0.

\bibitem{Barker2003}
M.~Barker, W.~Rayens, \href{http://dx.doi.org/10.1002/cem.785}{Partial least
  squares for discrimination}, Journal of Chemometrics 17~(3) (2003) 166--173.

\bibitem{Brown2001}
L.~Brown, T.~Cai, A.~DasGupta,
  \href{http://www-stat.wharton.upenn.edu/~tcai/paper/html/Binomial-StatSci.html}{Interval
  Estimation for a Binomial Proportion}, Statistical Science 16 (2001)
  101--133.

\bibitem{Pires2008}
A.~M. Pires, C.~Amado, Interval Estimators for a Binomial Proportion:
  Comparison of Twenty Methods, Revstat -- Statistical Journal 6~(2) (2008)
  165--197.

\bibitem{binom}
S.~Dorai-Raj, \href{http://CRAN.R-project.org/package=binom}{binom: Binomial
  Confidence Intervals For Several Parameterizations}, R package version 1.0-5
  (2009).

\bibitem{JaynesProbabilityTheory}
E.~Jaynes, Probability theory : the logic of science, Cambridge University
  Press, Cambridge, UK New York, NY, 2003.

\bibitem{Hmisc}
{F. E. Harrell Jr with contributions from many other users.},
  \href{http://CRAN.R-project.org/package=Hmisc}{Hmisc: Harrell Miscellaneous},
  {R} package v. 3.9-3 (2012).

\bibitem{Fleiss1980}
J.~L. Fleiss, A.~Tytun, H.~K. Ury, A Simple Approximation for Calculating
  Sample Sizes for Comparing Independent Proportions, Biometrics 36~(2) (1980)
  343--346.
\newblock \href {http://dx.doi.org/10.2307/2529990}
  {\path{doi:10.2307/2529990}}.

\bibitem{Vorburger2006}
M.~Vorburger, B.~Munoz, Simple Power Calculations: How Do We Know We Are Doing
  Them the Right Way?, in: Proceedings of the Survey Research Methods Section,
  American Statistical Association, 2006, pp. 3809--3812.

\bibitem{StatisticalMethodsForRatesAndProportions-Fleiss}
B.~L. Joseph L.~Fleiss, M.~C. Paik, Statistical Methods for Rates and
  Proportions, 3rd Edition, Wiley-Interscience, New Jersey, 2003.

\bibitem{ggplot2}
H.~Wickham, \href{http://had.co.nz/ggplot2/book}{ggplot2: elegant graphics for
  data analysis}, Springer New York, 2009.

\end{thebibliography}
